\documentstyle[mnextra]{mn}

\input{epsf}

\def\mincir{\,\raise -2.truept\hbox{\rlap{\hbox{$\sim$}}\raise5.truept
\hbox{$<$}\ }}
\def\Msun{M_{\odot}}
\def\magcir{\raise -2.truept\hbox{\rlap{\hbox{$\sim$}}\raise5.truept
\hbox{$>$}\ }}
\def\minmag{\raise-2.truept\hbox{\rlap{\hbox{$<$}}\raise 6.truept\hbox
{$>$}\ }}
\def\etal{{\it et al} \ }
\def\kms{km s$^{-1}$}
\def\kmsmpc{km s$^{-1}$ Mpc$^{-1}$\ }

\def\gsim{ \lower .75ex \hbox{$\sim$} \llap{\raise .27ex \hbox{$>$}} }
\def\lsim{ \lower .75ex \hbox{$\sim$} \llap{\raise .27ex \hbox{$<$}} }

\def\simlt{\mathrel{\spose{\lower 3pt\hbox{$\mathchar"218$}}
     \raise 2.0pt\hbox{$\mathchar"13C$}}}
\def\simgt{\mathrel{\spose{\lower 3pt\hbox{$\mathchar"218$}}
'     \raise 2.0pt\hbox{$\mathchar"13E$}}}

 \def\kpc{\,{\rm kpc}}    
 \def\Mpc{\,{\rm Mpc}}    
\def\kms{\, {{\rm km}}\,{{\rm s}}^{-1} }

\def\L{$\Lambda$CDM}
\def\L37{$\Lambda$CDM$_{0.3}$}

\def\be{\begin{equation}}
\def\ee{\end{equation}}

\pagerange{1--7}    
\pubyear{1998}
\volume{0}
\nobrackets
\seceqnum	
\begin{document}

\title[Dark Matter Halos within Clusters]{Dark Matter Halos within Clusters}
\author[Ghigna \etal]{Sebastiano Ghigna, Ben Moore, Fabio
Governato\\
Department of Physics, University of Durham, Durham DH1 3LE, UK.\\ \\
{\LARGE George Lake, Thomas Quinn, and Joachim Stadel} \\
Department of Astronomy, University of Washington, Seattle, WA
98195, USA.\\
\smallskip 
Email:S.S.Ghigna@durham.ac.uk\\}

\date{December 1997}
 
\maketitle  

\begin{abstract} 

We examine 
the properties of dark matter halos within a rich galaxy
cluster using a high resolution simulation that captures the
cosmological context of a cold dark
matter universe.  The mass and force resolution permit the resolution
of 150 halos with circular velocities larger than 80 $\kms\ $
within the cluster's virial radius of 2 Mpc.  This enables an
unprecedented study of the statistical properties of a large sample of
dark matter halos evolving in a dense environment.  The
cumulative fraction of mass attached to these halos varies
from 0\% at 200 kpc, to 13\% at the virial radius.  Even at this 
resolution the overmerging problem persists;
halos that pass within 200 kpc of the
cluster center are tidally disrupted. Additional substructure is lost
at earlier epochs within the massive progenitor halos. The median
ratio of apocentric to pericentric radii is 6:1; the orbital 
distribution is close to isotropic, circular orbits are
rare, radial orbits are common.  The orbits of halos are unbiased with
respect to both position within the cluster and with the orbits of the
smooth dark matter background and no velocity bias is detected.  The
tidal radii of surviving halos are generally well-fit using the simple
analytic prediction applied to their orbital pericenters.
Halos within clusters have higher concentrations than
those in the field. Within the cluster, halo density
profiles can be modified by tidal forces and individual encounters with
other halos that cause significant mass loss \-- ``galaxy harassment''.
Mergers between halos do not occur inside  the clusters virial radius.

\end{abstract}


\section{Introduction}
\label{s:Intro}

Rich clusters of galaxies are
large cosmological laboratories that may provide unbiased mixtures of
the matter content of the Universe (White \etal 1993, Evrard 1997).
Clusters are prominent structures in the Universe;
their evolution can be followed with samples out to $z \sim 1$
(Rosati \etal 1998).   
They are the most massive virialised objects in the
Universe and are the most recent objects to form in hierarchical
formation models.   Their masses can be determined by several
independent methods (e.g.  Carlberg \etal 1997, Cen 1997, Wu \&
Fang 1997, Smail \etal 1997). 

The evolution of the mass function of clusters is sensitive to key
cosmological parameters (e.g. Bahcall, Fan \& Cen 1997, Bartelmann
\etal 1997, Borgani \etal 1997, Carlberg \etal 1996, Eke, Cole \&
Frenk 1997, Wilson, Cole \& Frenk 1996).
This evolution can be calculated by using either
analytic methods (Press \& Schechter 1974) or cosmological N-body
simulations (Eke \etal 1997). The weakness of analytic methods is their
inability to follow halos that accrete into larger systems.  In the
past, numerical simulations have shared this problem.  To sample a
large volume, the poor resolution within virialised systems leads to
soft, diffuse halos that are rapidly dissolved by tidal forces. This
is the classic over-merging problem (White \etal 1987) that has 
lead to problems
when comparing the mass distribution within dark matter simulations to
the observed properties of galaxies (Summers \etal 1995).

We were determined to perform simulations that resolved
the distribution and evolution of galaxy halos within
clusters.  There are many questions that we will address with these
simulations: What is the orbital distribution of the galaxies within
clusters and are they biased in any way?  What is the extent of
galactic dark halos within clusters and how much of the cluster's mass
distribution is attached to galaxies?  How do these properties evolve
with time and within different cosmological models?  Are the density
profiles of isolated ``field'' halos similar to the halos that form
within the environment of a rich cluster?  How does the cluster
environment modify the internal structure of halos?  The frequency of
mergers between halos within the cluster environment and the heating
rate from halo--halo encounters are questions of importance for
studies of the morphological evolution of clusters.  These are amongst
the many questions that have remained unanswered due to the
over-merging problem.

Some have suggested that the only way to avoid merging is to follow
the evolution of the baryons---even if they are only $\sim 5\%$ of the mass
(Evrard \etal 1994).  This assertion seems counter-intuitive; it's now clear
that mass and force resolution can overcome the overmerging problem in
dark matter simulations (cf. Moore, Katz \& Lake 1996, Klypin,
Gottl\"ober \& Kravtsov 1997,  Brainerd, Goldberg  \& Villumsen 1997,
 Moore \etal 1997).
With fast parallel computers and highly tuned algorithms, our
simulations have hundreds of surviving halos within the virial radius
of a rich cluster (Moore \etal 1997).  Increased mass and force
resolution lead to higher central densities in galactic halos,
enabling greater survival within a cluster.  For the first time we can
compare the mass distribution with the galaxy distribution in a rich
cluster.

The plan of this paper is as follows: In \S 2, we describe the N-body
simulation, techniques and parameters.  In \S~\ref{s:iden} we
describe two algorithms to identify ``halos within halos''.  After
creating a catalogue of halos, \S~\ref{s:dist} turns to results
on their global properties; sizes, masses, radial distribution,
orbital properties and merger histories.  \S~\ref{s:pint}
discusses the internal properties of halos: their density profiles,
correlations between structure parameters and global parameters and
the evolution of these quantities with time.  We conclude in
\S~\ref{s:conc}.

\section{The $N$--body simulation}
\label{s:nbody}

One of the goals of performing cosmological numerical simulations is
to compare the distribution and bulk properties of dark
matter with the distribution and properties of the observable 
galaxies.  A direct comparison has never been possible, 
since structure in high density
regions has been quickly erased as a consequence of numerical resolution
-- the overmerging problem.  ``Galaxies'' are typically selected from 
the mass distribution in a cosmological simulation
using a biased sample of dark matter particles. Previous
studies of cluster substructure has been limited to using ``galaxy
tracers'' (Carlberg 1991, Summers \etal 1995), such as following the
most bound particle of a halo before it becomes disrupted by the tidal
field of the system.  These hueristics enable the use of low resolution
simulations, but their validity is anyone's guess
at this point (Summers \etal 1995).

There are now several codes
that are able to simulate a gaseous component.  Although these codes are
invaluable for many astrophysical problems; the original motivation
behind these techniques was the hope of resolving galaxies
in a cosmological context, thus solving the overmerging
problem.  They hoped to form galaxies and preserve them  
by increasing the central
densities owing to gas dynamics.  In turn, 
the halos would be more robust to disruption.
Indeed, SPH simulations of individual clusters, do give
rise to a set of galaxy tracers that resemble a ``real'' cluster.
However, the mass resolution in the dark matter component is not
sufficient to resolve the dark halos of the galaxy tracers,
typically one is left with a cold gas blob orbiting within the smooth
cluster background (Frenk \etal 1996).  There are 
a variety of pathologies that arise if one uses too few particles
to simulate a large dynamic range in scales.  Our simulations are
designed to resolve scales of 5 kpc using dark matter.  This would
be a minimal resolution to simulate galaxy formation with
gas dynamics.  One needs a fiducial dark matter simulation at
high resolution to see differences owing to gas
dynamics.  One must insure that these differences are sensible
as an external check.  With the gas representing $\lsim$10\%
of the mass, gross changes in numbers and orbits of galaxies would
be surprising.

Our aim is to achieve very high spatial and mass resolution within a
rich virialised cluster drawn from a ``fair volume" of
100 ${\rm Mpc}^3$ in a standard CDM Universe.  
In such a volume, there are several rich
clusters, and none dominate the environment in an undesirable way.
Any simulation method limits the number of particles that can be
invested in a single simulation. Previous simulations of clusters suffered 
limitations due to the small volume used, forcing the run to stop
at $z \sim 1$,  (Evrard, Summers \& Davis 1994),
or or the use of vacuum boundary conditions outside the cluster 
(Carlberg 1991).


With current technology, we can perform a single large simulation of
$\sim 10^8$ particles or try to tackle a number of different problems
using simulations with $\sim 10^7$ particles.  If we simulate our
``fair volume" at uniform resolution, 
there will be $\sim 10^4$ particles within the virial
radius of a cluster, a resolution that is insufficient to resolve
substructure.  Previous attempts to resolve the inner structures of
cluster halos using $\sim 10^5$ particles 
failed to resolve more than a handful
of satellite halos (Carlberg 1991, Carlberg \& Dubinski 1991 , Tormen 1997, Tormen {\it et al}).  To achieve
higher resolution within an individual cluster we initially perform a
simulation of a large volume of a CDM universe as described above,
normalised such that $\sigma_8=0.7$ and the shape parameter
$\Gamma=0.5$ (H=50\,\kmsmpc\ is adopted throughout).

We used a nesting scheme that we call ``volume renormalization" to
achieve higher resolution within a region of greater interest.  This
technique 
has been used to probe quasar formation at high redshift (Katz \etal
1994) and to follow the density profiles of halos in a cosmological
context (Navarro \etal 1996).  We generate initial conditions (ICs)
for the volume at two resolutions, one that places $\sim 10^7$
particles within the entire volume and one such that there would be
$\lsim 10^6$ particles in the targeted cluster. We run the lower
resolution model and select a virialised cluster at $z = 0$.  The
particles within about twice the virial radius of the cluster in the
final state are traced back to their locations in the ICs.  Within
this region, we use the higher resolution ICs.  Beyond this high
resolution region the mass resolution is decreased in a series of
shells by combining particles in the high resolution ICs at their
center of masses.  In this way, the external tidal field is modelled
correctly.  The starting redshift in the high resolution run is
increased to z=69 such that the perturbations in the smoothed density
field of the high resolution region obey the constraint
$\delta\rho/\rho \lsim 0.1$.  We then re-run the simulation to
the present epoch.

We use a new high performance parallel treecode ``PKDGRAV'' to evolve
the particle distribution.  PKDGRAV (Stadel \etal, in preparation) has
accurate periodic boundaries and an open ended variable timestep
criteria based upon the local acceleration (Quinn \etal 1997).  The
code uses a spline softening length such that the force is completely
Newtonian at twice our quoted softening lengths.  In terms of where
the force is 50\% of the Newtonian force, the equivalent Plummer
softening length would be 0.67 times the spline softening length.
In the high resolution region, our particle mass is
$8.6\times10^8M_\odot$.  We perform two runs with 10 kpc (RUN1) and 5
kpc (RUN2) softening lengths. The final virial radius of the cluster
is $\sim 2$ Mpc and the mass is $4.6\times 10^{14}M_\odot$ so that we
have approximately 600,000 particles within a sphere of overdensity
200. (Note that the cluster that is analysed here is the ``Virgo''
cluster from Moore \etal 1997.)

\section{Cluster substructure and halo identification}
\label{s:iden}

\subsection{The density distribution in the cluster}

\begin{figure*}
\begin{picture}(400, 650)  
\put(0, 470)               
{\epsfxsize=12.truecm \epsfysize=7.truecm 
\epsfbox[40 300 600 720]{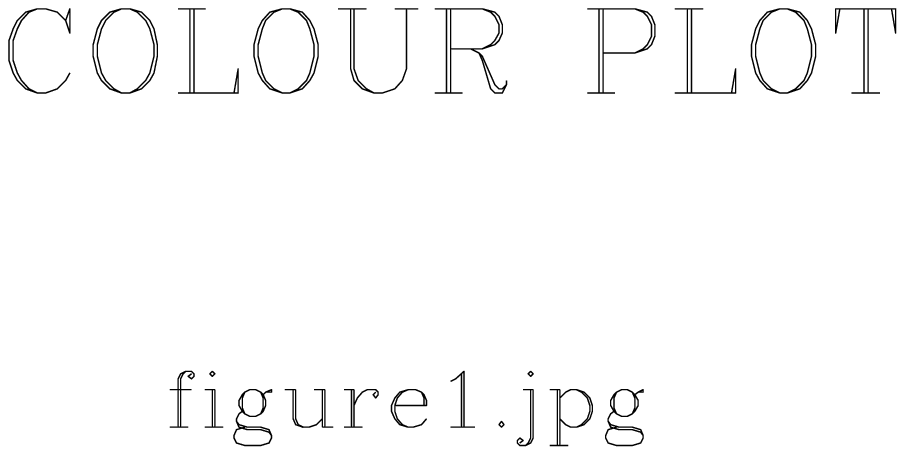}}
%
\put(0, 140)               
{\epsfxsize=11.truecm \epsfysize=7.truecm       
\epsfbox[40 300 600 720]{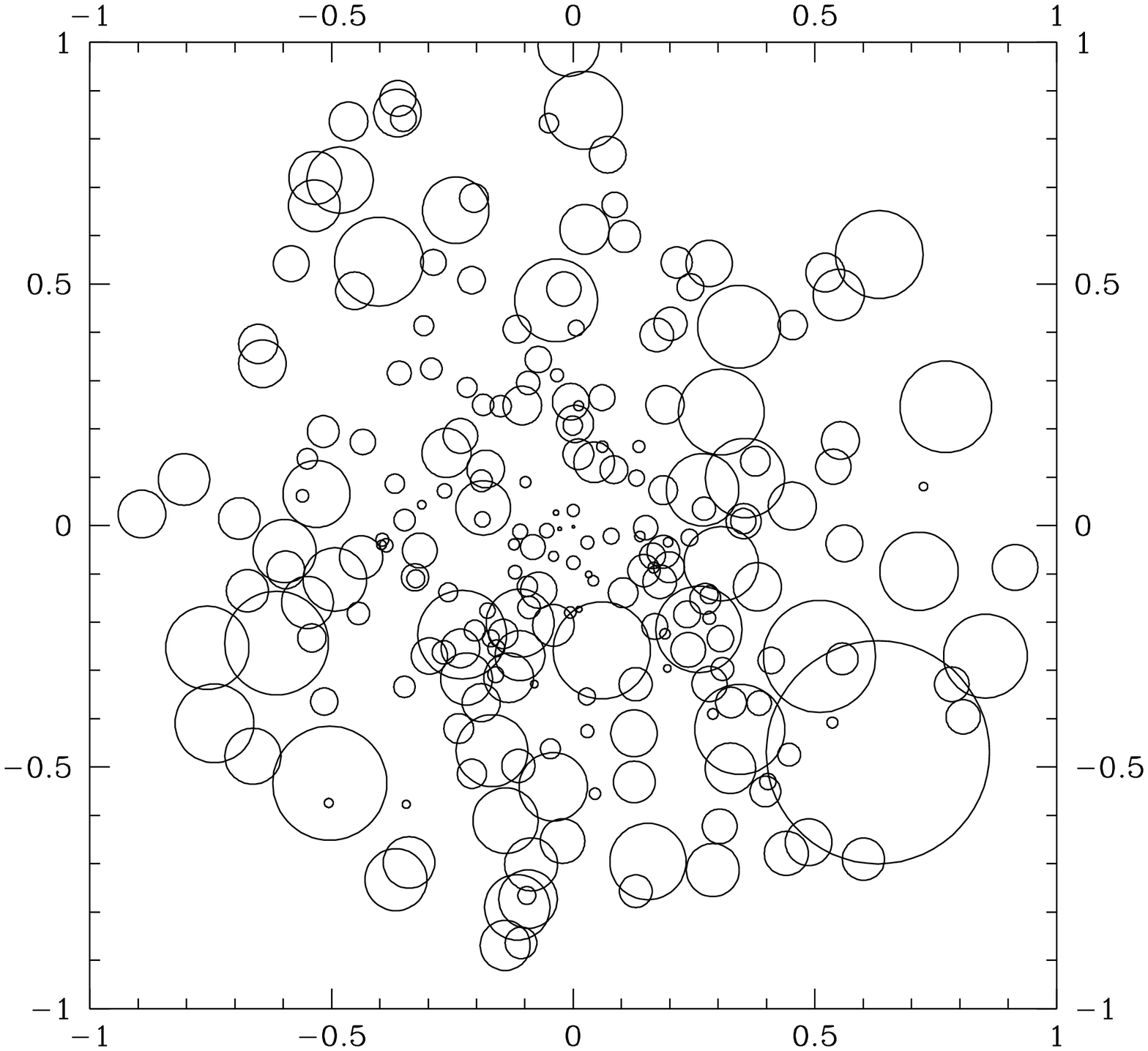}}
\end{picture}
\caption{Density map (colour plot) and circle plot of
the halo radii within the cluster's virial radius 
(taken here as the length unit) at $z=0$.} 
\label{f:clustview}
\end{figure*}

In past work, halos in dissipationless $N$--body simulations have
usually spontaneously dissolved when entering clusters.  Two physical
effects conspire with the finite numerical resolution to erase dark
matter halos in clusters (Moore, Katz \& Lake 1996).  Halos are heated
by cluster tides and halo--halo encounters, thus losing mass as they
move into the potential well. When the tidal radius approaches $\sim
3$ times their ``core radii" (owing to either a density plateau or
gravitational softening), they dissolve.  Hence, it takes very high
resolution to retain dark matter substructures at a distance 100-200
kpc from the cluster's center.  Our numerical parameters were chosen
so that halos would survive at these scales.  The wealth of halos
retained in our simulated cluster is visible in 
Figure~\ref{f:clustview}

The upper panel is a map of the
density distribution in a box of size $R_{200}$ 
(see \S 3.2 for a precise definition), centered on the cluster
and projected onto a plane.
 Each particle is colour coded 
according to the logarithm of the local density
(defined using an SPH smoothing kernel over 64 particles
in a code called {\sl SMOOTH} [Stadel \& Quinn 1997,
http ref: http://www--hpcc.astro.washington.edu/tools]).
Only regions with density
contrast $\delta>30$ are shown.  The cluster boundaries, set
at $R_{200}$, correspond to the contours of the central blue region.
Much of the mass inside $R_{200}$ lies in the dark matter
halos that we will analyze here. Their projected distribution is shown
as a `circle plot' in the lower panel of Figure~\ref{f:clustview}. 
The
radius of each circle is the halo `tidal radius' (Sec.~3.3) in units
of $R_{200}$. Note that halos of similar central densities (similar
colour in the density map) may have largely different radii depending on
their distances from the cluster's center. 
It is remarkable that substructure halos cover such a large
fraction of the projected cluster's area.

\subsection{Cluster properties and evolution}

\begin{figure}
\centering
\epsfxsize=\hsize\epsffile{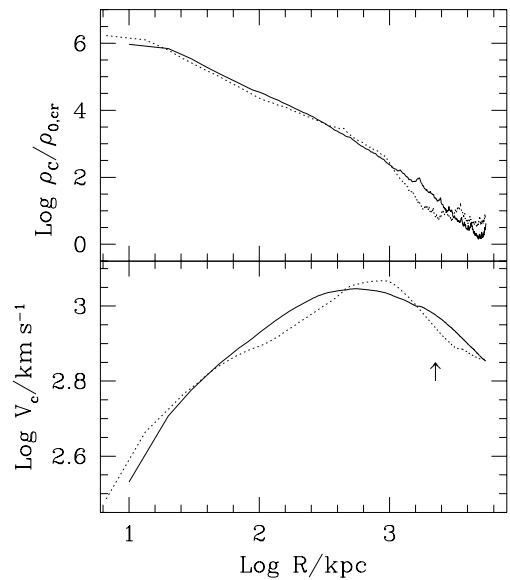}
\caption{The density (upper panel) and circular velocity (lower panel)
profiles for the cluster. The solid curves are at the final epoch and
the dotted curves show the cluster at a redshift $z=0.5$.  The density
is measured in units of the present critical density $\rho_{0,cr}$;
$R$ is the physical distance to the cluster center, and was set at
the location of the most bound particle. The arrow marks the value of
the cluster virial radius $R_{200}$ at $z=0$.}
\label{f:clustprfs}
\end{figure}
We define the cluster center as the position of its most bound
particle.  This particle 
is within a softening length of the center
of the most massive halo found by SKID.  The density profile of the
RUN1 cluster calculated in spherical shells
is shown in the upper panel of Figure~\ref{f:clustprfs}
(the solid line is for $z=0$ and
dotted line for $z=0.5$).  The cluster forms at $z\sim 0.8$ from the
mergers of many halos along a filamentary structure and, at $z=0.5$,
it has not yet virialised since it still has quite a lumpy structure
but the global density profile is roughly similar to that measured 5
Gyrs later at z=0. (We shall compare properties of the substructure
identified at both epochs.)

The lower panel of this figure shows the circular velocity profile
$V_{c}(R)\equiv (GM(R)/R)^{1/2}$, where $M(R)$ is the mass within $R$.
The virial radius of the cluster is defined as the distance
$R_{200}$ for which the average density enclosed,
$\bar\rho_C(R_{200})$, is 200 times the cosmic density, $\rho_{cr}$;
we obtain $R_{200}=1.95\Mpc$ at $z=0$ and $1.2\Mpc$ at $z=0.5$.
The cluster is not spherical and has axial ratios that are roughly
2:1:1. (In the following we will always use units of kpc and $\kms$
for lengths and velocities, unless we explicitly state otherwise.) 

\begin{figure}
\centering
\epsfxsize=\hsize\epsffile{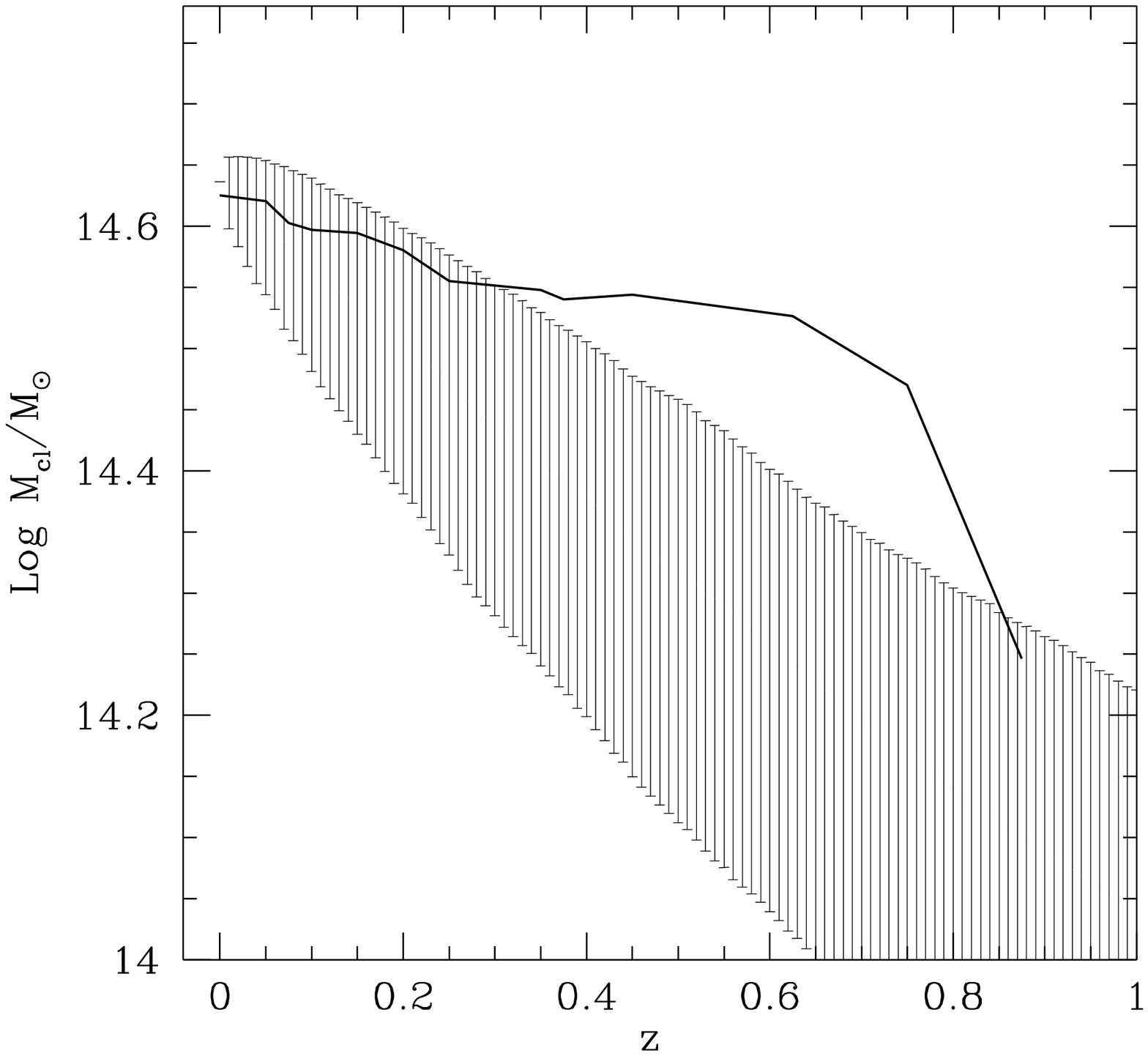}
\caption
{Growth of the cluster's mass $M_{cl}$ with redshift $z$.
$M_{cl}$ is the mass encompassed by the cluster's `virial' radius
at each $z$. The errorbars delimit the 1--$\sigma$ confidence interval
for the growth rate predicted using the Press-Schechter theory.}
\label{f:Mgrowth}
\end{figure}
Figure~\ref{f:Mgrowth} shows the growth of the cluster's mass with
redshift.  Defining the formation redshift of the cluster as that where
it has accreted half of its final mass, $z_{form}\sim 0.8$
for our cluster. This is slightly earlier than expected for an average
 cluster of this mass from  the Press--Schechter theory
 (1974; Lacey \& Cole 1993), where $z_{form}\sim0.5$
(as calculated from a routine kindly supplied by Paolo Tozzi).
This is not an unwelcome feature, since the cluster is in true virial
equilibrium at $z=0$.

\subsection{Halo identification}

Identifying DM halos in the high--density environment of the cluster
is a critical step (cfr. Klypin, Gottl\"ober \& Kravtsov 1997).  The
halos jump out visually, so while  it is relatively straightforward to
identify the halo centers, we must select only the bound particles to
characterize the halo.  We want to screen the cluster background
that's streaming through, but the substructure itself will be tidally
distorted and may have tidal tails of material that are loosely
attached to the halo.  Our group finding algorithm uses local density
maxima to find group centers and and then iteratively checks for
self--boundedness to define group membership. Each {\sl group} of
particles found belongs to an individual {\sl halo}.  The algorithm is
an improved version of DENMAX (ref), named SKID and is fully described
by Stadel \etal (1996, see http ref:
http://www--hpcc.astro.washington.edu/tools).  For each simulation we
adopt a linking length of $1.5l_{soft}$ and a minimum number of member
particles of 16, corresponding to a mass of $\simeq 1.35 \cdot
10^{10}\Msun$.  In general, we shall use halos with more than 16
particles when they are employed as tracers, but we shall adopt a
minimum number of 32 particles when their individual properties are
relevant.

The high resolution region that we analyze is roughly the turn-around
radius of the cluster or about twice the virial radius,
$R_{ta}\simeq 2 R_{200} \simeq  5\Mpc$. 
Within this radius, there are 495 and 522 halos for RUN1 and RUN2, and
208 and 227 halos respectively within $R_{200}$.  Changing $l_{soft}$
by a factor of two does not make much difference on a global scale,
but if we restrict ourselves to the inner parts of the cluster the
difference between the numbers of halos changes significantly: Inside
$R<R_{200}/2$, RUN2 has 91 halos compared to the 59 found in RUN1 and
RUN2 has twice as many within $R<R_{200}/4$ (30 instead of 16). The
innermost halos in RUN1 and RUN2 are at $\sim 200\kpc$ and $\sim
100\kpc$ respectively.  These differences reflect the 
softening length's effect on the halos' central densities that
determine their survival against tidal disruption (Moore
\etal 1997; see also the discussion related to Figure~\ref{f:rcore}
below).  Quality control of our halo finding algorithm was insured by
visually inspecting the density distribution inside $R_{200}$ to
verify that we neither missed nor created structures.

\begin{figure}
\centering
\epsfxsize=\hsize\epsffile{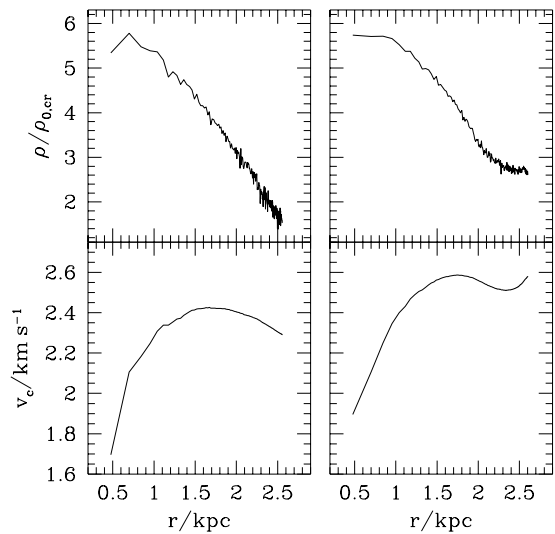}
\caption{Density profiles and circular velocity profiles are plotted
for two large dark matter halos extracted from the simulation. The
left panel is a "peripheral" halo that lies beyond the main cluster's
virial radius at $R=4.2$ Mpc.  The halo in the right panel lies within
the cluster at $r=1.2$ Mpc.  The radius $r$ measures the distance from
the center of each halo.  In the latter case, the radius at which the
high density background of cluster particles dominates the halo
particle distribution is apparent as a flattening of $\rho(r)$ and the
linear rise of $v_c(r)$ in the outer regions.
We denote the peak value of the circular velocity as $v_{peak}$ and
the radius at which it occurs as $r_{peak}$.}
\label{f:profiles}
\end{figure}
We use the output of SKID to determine the halo structural parameters. 
In particular, it estimates the extent of a halo using the distance to
the least bound particle.  However, the full 6--dimensional phase
space information is never available in the real Universe, therefore
we shall compare results from SKID using an ``observable'' quantity
for each halo.  For example, the (projected) mass distributions can be
determined using either weak lensing, the rotation curves of spirals
or the velocity dispersion profiles of ellipticals.
Motivated by gravitational lensing, one possibility is to define a 
halo's radius using
its density profile $\rho(r)$, where $r$ is the distance from a halo's center, 
and measuring the radius ($\equiv r_\rho$) at which 
the local density of the cluster background dominates and $\rho(r)$ flattens
(cfr. fig.~\ref{f:profiles}). Rotation curves or velocity 
dispersion profiles of isolated objects will eventually
decline with radius, but if a halo is embedded within a deeper
potential, at some radius its profile will turn
around and increase as the velocity dispersion of the 
cluster background starts dominating. 
(Figure 4 shows that the position of these inflexion points are essentially equal.)
We therefore combine these two definitions and use  
the inflexion point ($\equiv r_{vc}$) of the effective circular velocity 
$v_c=(G M(r)/r)^{1/2}$ as an alternative independent estimate of the extent
of a halo (cfr. fig.~\ref{f:profiles}). 

The circular velocity is less noisy than the density or velocity
dispersion and thereby more suitable for an
automated procedure.  Moreover, the inflexion point of $v_c(r)$ can be
easily detected by searching for a minimum, without any
knowledge of the background density (as it would be necessary, if we
were to implement an overdensity criterium).  The radius $r_{vc}$ can
either overestimate $r_\rho$, for steeply declining velocity profiles,
or underestimate it, for profiles close to isothermal.  If the halos
are described by the NFW model (Navarro, Frenk \& White 1996; see also
Section~5), the former condition applies to halos with $v_c \sim
50\kms$ and the latter to halos with $v_c \magcir 200\kms$ for
background densities $\sim$ 300 times the cosmic average (as we will
see however, the profiles of tidally `stripped' halos decline with $r$
more steeply than an NFW profile).  In our case, for small halos, the
difference can be at most $\sim 10$\%; as for large halos, $r_{vc}$
differs from $r_\rho$ by $\mincir$ 20\% in about $1/3$ of the halos in
our sample with $v_c\magcir 120\kms$. These differences do not
significantly affect our results.  With this definition, there is a
contribution to the mass encompassed by a halo from the smooth
background of order 20\%, which we subtract from the quoted halo
masses.

For each SKID halo, we calculate $\rho(r)$ and $v_c(r)$ using equally
spaced bins of 2 kpc, such that the number of particles in each bin
is nearly equal.
The departure of $\rho(r)$ from isothermality is betrayed by
a peak of $v_c(r)$, $v_{pk}$ occurring at $r_{pk}$.  The catalog
values of $r_{pk}$ and $v_{pk}$ are estimated by fitting a cubic
spline to $v_c(r)$.  The left panels of Figure~\ref{f:profiles} show a
sample $\rho(r)/\rho_{0,cr}$ and $v_c(r)$ curves for a large {\sl
peripheral halo} at a distance of $R=4.2\Mpc$ from the center of RUN1
(hereafter {\sl cluster halos} are those within $R_{200}$, and {\sl
peripheral halos} those between $R_{200}$ and $R_{ta}$).

Halos beyond the cluster boundary can be easily characterized by their
`virial' radii and masses $M_{200}\equiv M(r_{200})$.  For isothermal
spheres the circular velocity at $r_{200}$ is $v_{200} =
(r_{200}/\kpc)h\kms$ to within a few percent.  For the halo mentioned
above, $r_{200}\simeq400\kpc$ ($M_{200}=3.4\cdot 10^{12} \Msun$).  The
right panels of Figure~\ref{f:profiles} show $\rho(r)/\rho_{0,cr}$ and
$v_c(r)$ for a tidally limited massive cluster halo at $R=1.2\Mpc$,
where $\rho_{bkg}\simeq300\rho_{0,cr}$.  The halo radius $r_{halo}$ is
either the virial radius, $r_{200}$ or the {\sl tidal radius} $r_{tid}$,
as appropriate.  Similarly, the halo mass is defined as
$M_{halo}\equiv M(r_{halo})=v_c(r_{halo})^2r_{halo}/G$.

Force softening and the finite mass resolution introduce halo {\sl
cores} that are visible in Figure~\ref{f:profiles} as a flattening of
$\rho(r)$ at $r\mincir 10\kpc$.  We define a {\sl core radius}
$r_{core}$ for each halo as the radius where $v_c$ has risen to 70\%
of $v_{peak}$.

Once the halo positions and structural parameters are known, we can
start to address the questions raised in the Introduction.  We begin
by examining the global properties of the halo distribution, with a
preliminary discussion of numerical resolution effects.

\section{Properties of the halo distribution}
\label{s:dist}

\subsection{Numerical resolution effects}
\label{sub:num}

\begin{figure}
\centering
\epsfxsize=\hsize\epsffile{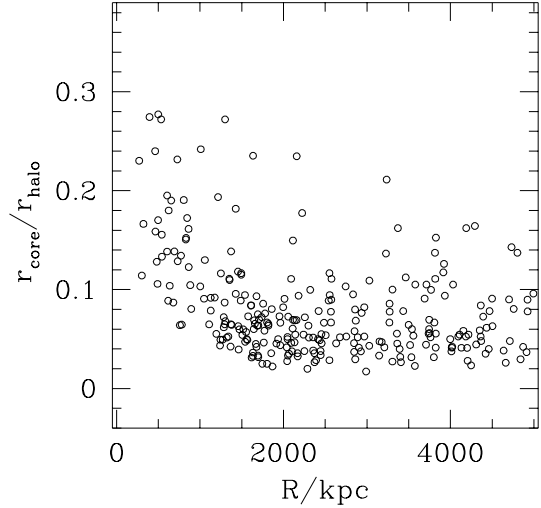}
\caption{For every halo within $R=5$ Mpc we plot the ratio of halo 
core radii and outer radii vs. clustercentric position. Halos contain
at least 32 particles.
}
\label{f:rcore}
\end{figure}
Halo in clusters are ``harassed" by encounters with other halos
combined with strong global tides.  This adds internal energy and
leads to mass loss.  Using a static cluster potential, Moore, Katz \&
Lake (1996) found that halos rapidly disrupt when $r_{tidal}<3$--$4
r_{core}$.  Our fully self-consistent simulations verify this.
Figure~\ref{f:rcore} shows the ratios $r_{core}/r_{halo}$ $vs$ $R$ for
our halo set using RUN1.  This ratio increases towards the cluster
center and no halos exist with $r_{core}/r_{halo}\magcir 0.3$.
(Similar results hold for RUN2.)  Clearly, the softening will set a
floor to the core radius, but in general we find that $r_{core}\sim
l_{soft}$, and correspondingly, that halos smaller than $\sim
3l_{soft}$ have all dissolved.  The lower boundary for $r_{halo}$ in
RUN1 and RUN2 is indeed $\sim$30 and $\sim$15 kpc respectively.

\begin{figure}
\centering
\epsfxsize=\hsize\epsffile{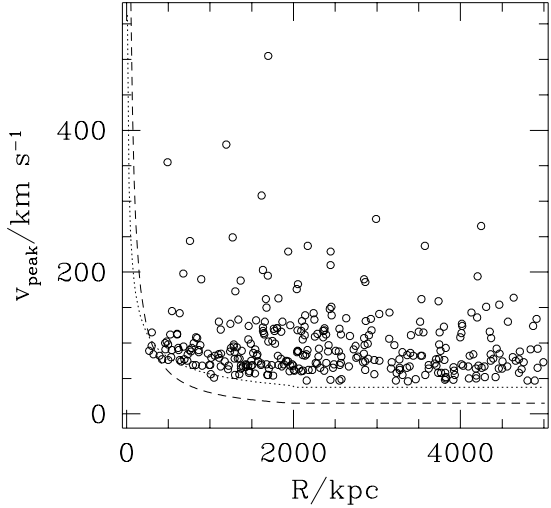}
\caption{The peak circular velocities of halos, $v_{peak}$, is plotted against
their clustercentric position $R$ at $z=0$. The lines give the
expected limiting $R$ at which halos of given $v_{peak}$ can be resolved
because of the finite spatial resolution (dashed curve) and mass resolution 
(dotted line).}
\label{f:vpeak}
\end{figure}
Numerical limitations can also be seen in Figure~\ref{f:vpeak}, a plot
of $v_{peak}$ versus $R$ for RUN1.  There is no bias apparent in this
plot, halos of all circular velocity exist over a wide range in $R$.
We show two curves that exclude regions of space owing to our
numerical resolution: $(i)$ defining a minimum particle number to
identify halos sets a lower mass limit, $M_{lim} =1.35\cdot
10^{10}\Msun$, $(ii)$ $r_{tidal}$ is correlated with $R$ and halos
dissolve when $r_{tidal} \magcir 3 r_{core} \simeq 3 l_{soft}$.
Approximating the halos as isothermal spheres that are tidally
stripped like layers off an onion, the mass and circular velocity are
related as: $M_{halo}\equiv M_{200} = G v_c^2 r_{200}$ for unstripped
peripheral halos and $M_{halo}\simeq G v_c^2 r_{tid}$.  for
``stripped'' cluster halos.  The tidal radius 
obtained from $v_c$ through $r_{tid}\simeq R v_c/V_c$ ($V_c\simeq
1000\kms$ is the circular velocity for the cluster) in the
approximation that a halo at $R$ is tidally truncated ``locally'',
{\it i.e.} according to the value of $\bar\rho_C$ at $R$ (this is a
limiting case as their pericentric radius can only be smaller).
Applying condition $(i)$ leads to the dotted line in
Figure~\ref{f:vpeak}; criterion $(ii)$ leads to the dashed line (we
set $v_c={\mbox{\rm const}}=v_{peak}$).  If tides due to the cluster
potential are the only cause of halo disruption, our sample should be
complete for halos with $v_{peak}\magcir 80\kms$ and pericenters that
have always been greater than $R_{lim}\sim 250\kpc$.

\subsection{Spatial distribution of halos}

\begin{figure}
\centering
\epsfxsize=\hsize\epsffile{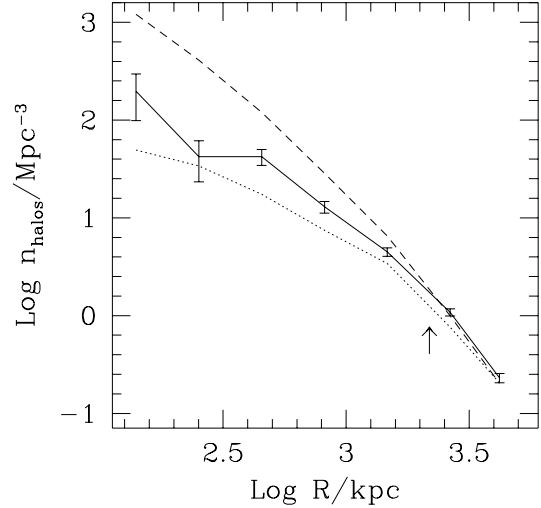}
\caption{The number density of halos with $v_{peak}\ge 80\kms$ as a
function of $R$, for RUN1 (dotted line) and RUN2 (solid line); the
former has a spline softening length of 10 kpc, twice that of the
latter.  The dashed line is the particle density scaled in such a
way that, in the distance interval $[R_{200},R_{ta}]$, the {\sl
scaled} number of particles equals the number of halos. The errorbars
are 1--$\sigma$ Poissonian errors from the counts in each bin; the
arrow marks the value of $R_{200}$.
}
\label{f:ndenslg}
\end{figure}
The number density of halos, $n_{halos}$, is plotted against $R$ in
Figure~\ref{f:ndenslg}, for the two runs.  The dashed line refers to
the particle density, $n_{part}$, normalised to 
the halo density in the interval $[R_{200}, R_{ta}]$.

Beyond the cluster's virial radius the curves for halos and particles
have similar slopes. Within the cluster's virial radius we see that
the halo distribution is ``antibiased'', {\it i.e.} less
concentrated, with respect to the mass distribution.
The halo number density profile is consistent, for $R> 1\Mpc$, with
the average galaxy number density profile derived by Carlberg \etal
(1997) for the clusters of the CNOC survey.  However, the average
cluster mass profile that they derive from the same data is much
shallower than that of the relatively small cluster analyzed here.

If we regard the scaled particle distribution as the ``asymptotic''
halo distribution in the case of no bias and infinite resolution (and
sufficiently small physical cores), then we can estimate the numbers
of missing halos by integrating the difference over $R$.  For RUN2 we
find that $\sim 240$ halos are ``missing'', about 50\% of the halos
with $v_{peak}>80\kms$.   The
systematic difference between RUN1 and RUN2 makes tidal disruption of
``softened'' halos the most likely cause of this bias.  
However, the destruction of halos may not be due to
numerical resolution alone.  For example, binary mergers between halos
of similar mass will lead to a single halo with no memory of its
history.  This halo anti-bias may be overcome with higher resolution
simulations that accurately resolve the structure of the smallest halos.

\subsection{Distribution of halo radii}

The extent of dark matter halos attached to galaxies in clusters has
become directly observable via observations of gravitational lensing
(Geiger \& Bartelmann 1997, Natarajan \etal 1997).  Here we can make
some predictions for future surveys that will constrain the extent of
halos as a function of position from the cluster center.

The projected distribution of the halos within the cluster's boundaries
was shown in the lower panel of Fig.~\ref{f:clustview}. 
There is a clear decrease of halo
sizes near the cluster's center even in this projected plot
(there is
little difference if we include halos up to twice $R_{200}$).

\begin{figure}            
\begin{picture}(300, 430)  
\put(0, 280)
{\epsfxsize=8.0 truecm \epsfysize=6.truecm 
\epsfbox[40 300 600 720]{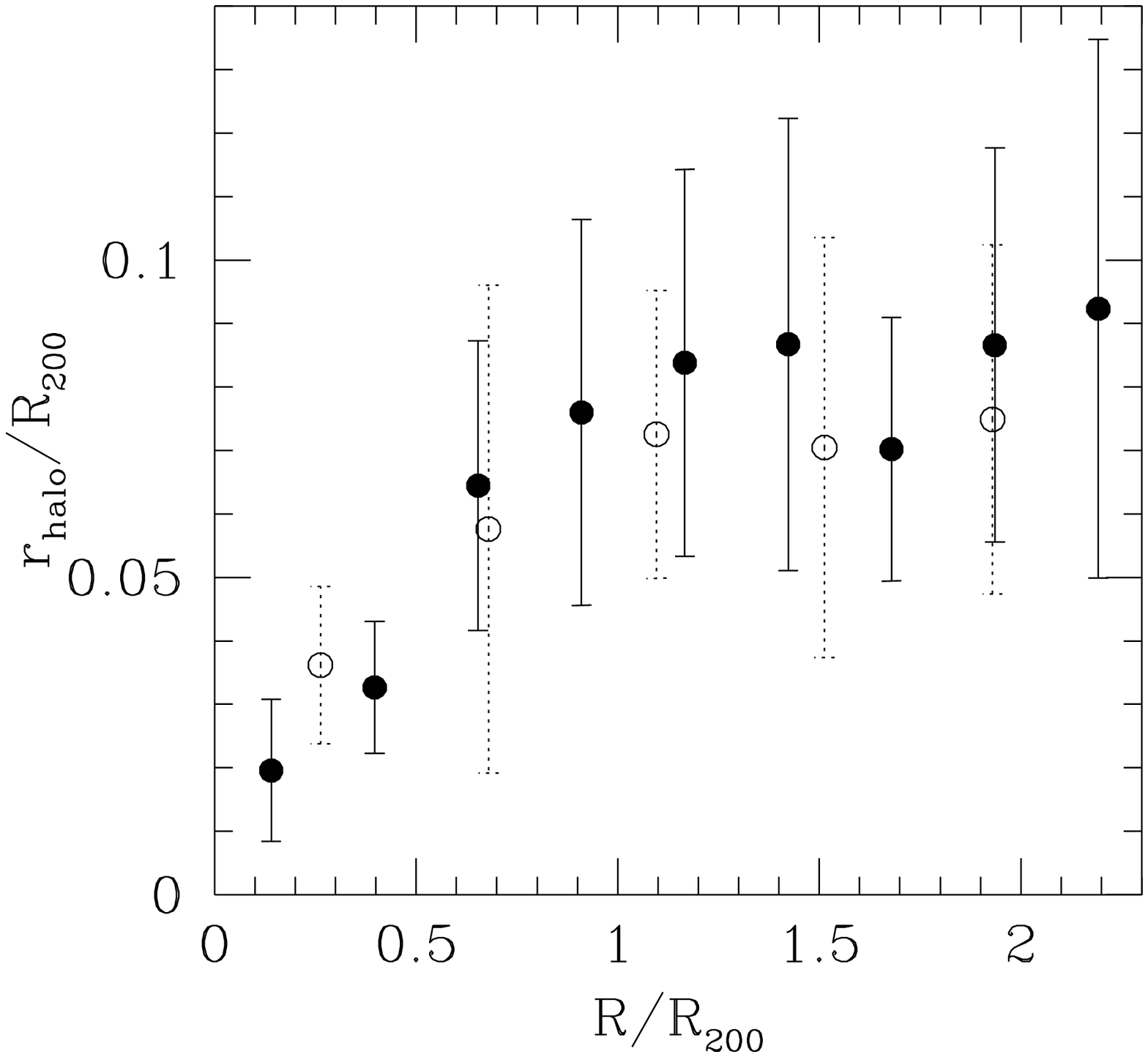}}
\put(0, 70)
{\epsfxsize=8.0truecm \epsfysize=6.truecm 
\epsfbox[40 300 600 720]{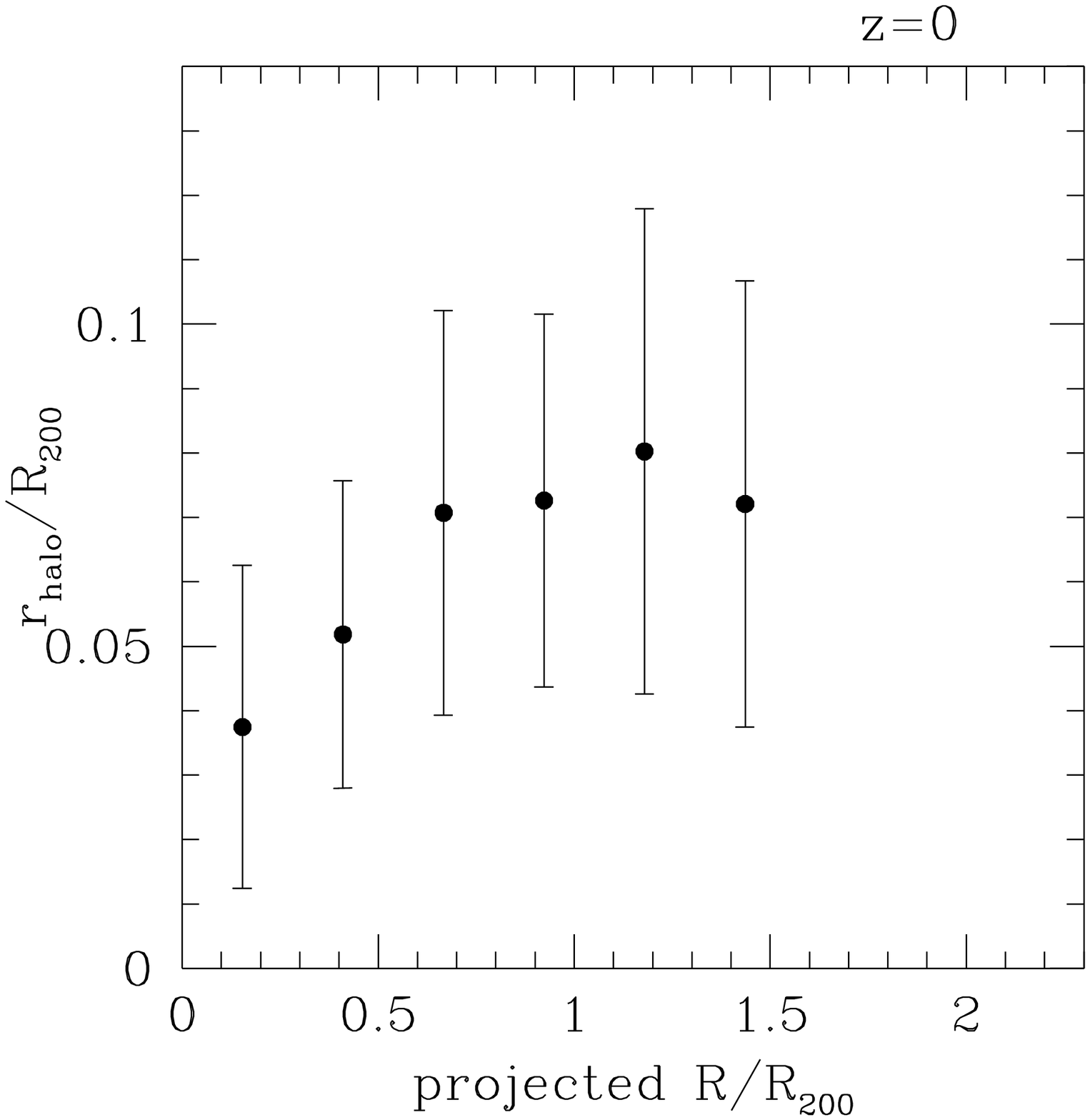}}
\end{picture}
\caption{In the upper panel, the average value of halo radii plotted against
cluster-centric position at redshift $z=0$ (solid symbols) and $z=0.5$
(dotted), for halos with $v_{peak}\ge 80\kms$.  The lower panel
shows the distribution against projected distance at $z=0$.
At each $z$, the unit is the cluster's virial radius at that epoch.
The errorbars give the dispersion about the
average.  Note that the most massive cluster halo at $z=0$ has
been excluded from the average; it has a radius of 470 kpc and is
located at $R=0.9 R_{200}$.}
\label{f:rhal-R}
\end{figure}

Figure~\ref{f:rhal-R} shows the average value of $r_{halo}/R_{200}$ as
a function of clustercentric position, $R/R_{200}$,
at redshifts z=0 and z=0.5.  
We use $R_{200}$ at each $z$ as the length unit 
to highlight the self--similarity
in the evolution of the cluster substructure. 
Halo radii clearly decrease as we move towards
the cluster center, 
but the trend is hard to detect at $z=0.5$
because the cluster 
has accreted only relatively few halos (of $v_{peak}>80\kms$), 
has a quite anisotropic mass distribution and 
tides have been efficient only in its very center. 
The mean size of halos beyond $R_{200}$ is
$\magcir 8$\% of $R_{200}$   
and drops approximately linearly to
zero as we move from the virial radius to the cluster center. 
(In Fig.~\ref{f:rhal-R},
 halo radii are those measured from the circular
velocity profiles; the results using SKID radii are very similar.)
As shown in the lower panel of this Figure, projection effects 
weaken but do not erase the trend of halo radii with $R$ (here
we have included halos up to $1.5 R_{200}$, but it makes little
difference changing the limiting $R$ between once and twice $R_{200}$). 
Natarajan et al. (1997) do not detect a dependence on $R$, but
presently, the observational uncertainties are large and 
the radial range of the data is limited.

\begin{figure}             
\begin{picture}(300, 430)  
\put(0, 280)
 {\epsfxsize=8.truecm \epsfysize=6.truecm 
\epsfbox[40 300 600 720]{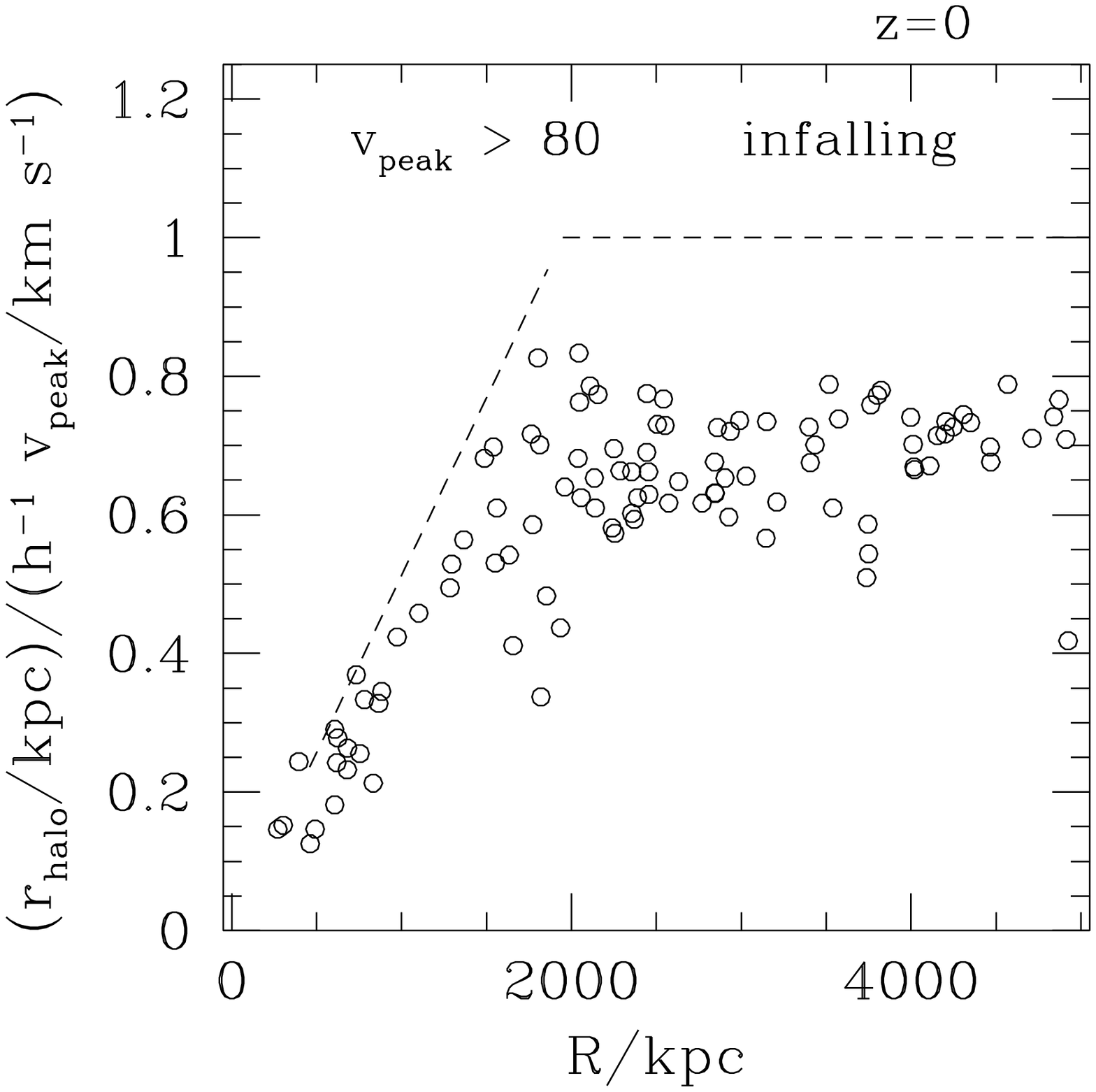}}
\put(0, 70)
{\epsfxsize=8.truecm \epsfysize=6.truecm 
\epsfbox[40 300 600 720]{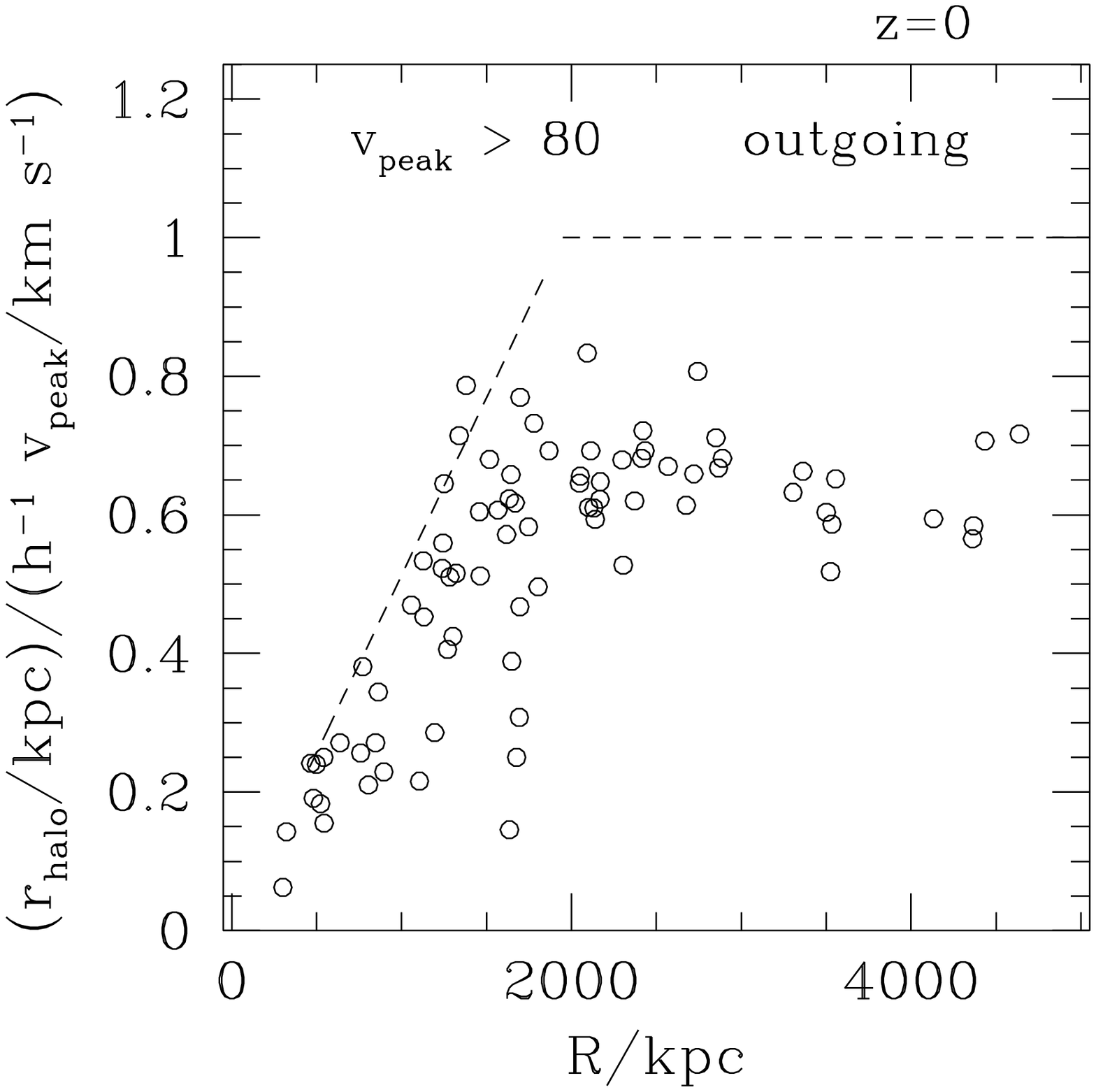}}
\end{picture}
\caption{The figure highlights the effects of tidal stripping on halo
sizes at varying distance from the cluster's center (halos with
$v_{peak}>80\kms$). For our halo sample,
the ratio $r_{halo}h/v_{peak}$, with $h=h_0(1+z)^{1.5}$ and units of
kpc and $\kms$, is approximately 70\% of the value expected for a
purely isothermal mass distribution. The dashed line gives the
expected behaviour for isothermal halos being instantaneously stripped
whilst falling into the deeper potential of the cluster. The upper and lower
panels are for inwards and outwards moving halos respectively. See the text
for further details.}
\label{f:rhalvpk}
\end{figure}

Halos of different $v_{peak}$ obviously have intrinsically different
sizes.  We can interpret Figure~\ref{f:rhal-R} and account for this
intrinsic scatter using our ``locally stripped isothermal onion
approximation" (with $v_c\equiv v_{peak}$; see Section~4.1) and
considering the ratios $\eta\equiv r_{halo}/(h^{-1} v_{peak})$.  When
halos form in isolation, $r_{halo} \equiv r_{200}\approx h^{-1}v_c$
and $\eta \simeq \mbox{\rm const} \simeq 1\cdot \kpc/\kms$.  Within
the cluster, $r_{halo}\equiv r_{tid}\simeq R v_c/V_c \simeq (h^{-1}
v_c) R/R_{200})$ and $\eta(R)\simeq R/R_{200}$.  If halos have gone
past pericenter $r_{peri}$ and their radii $r_{tid}$ are determined by
$\bar\rho_C(r_{peri})$, then $\eta$ will be smaller by a factor of
$r_{peri}/R$ (cfr. Fig.~\ref{f:rhalvpkScor} in Sec.~\ref{ss:orbits}).

In a fully virialized system there is no difference between the
motions away and towards the system's center. However, in a $\Omega=1$
cosmology, clusters never stop accreting material and we might expect some
different morphologies between ingoing and outgoing halos. Moreover, 
halos reaching apocenters beyond the cluster's boundaries might be 
heated by low--speed encounters with other halos and re--expand, partly
covering the effects of tides at the previous passage at pericenter.

The distribution of $\eta$ at $z=0$ is shown in Figure~\ref{f:rhalvpk}. 
It has a trend similar to that  
expected for locally truncated isothermal spheres (dashed line). 
The separation between the latter and the points in the ``periphery''
($R>2\Mpc$) is a measure of the departure of the actual density
profiles of peripheral halos from isothermal.
All outgoing halos must have necessarily passed their pericenters;
therefore, $\eta$ should depart from the dashed
line more than those for infalling halos and we do observe
evidence of this effect in the Figure. 
\begin{figure}
\begin{picture}(300, 430)  
\put(0, 280)
 {\epsfxsize=8.8truecm \epsfysize=6.truecm 
\epsfbox[40 300 600 720]{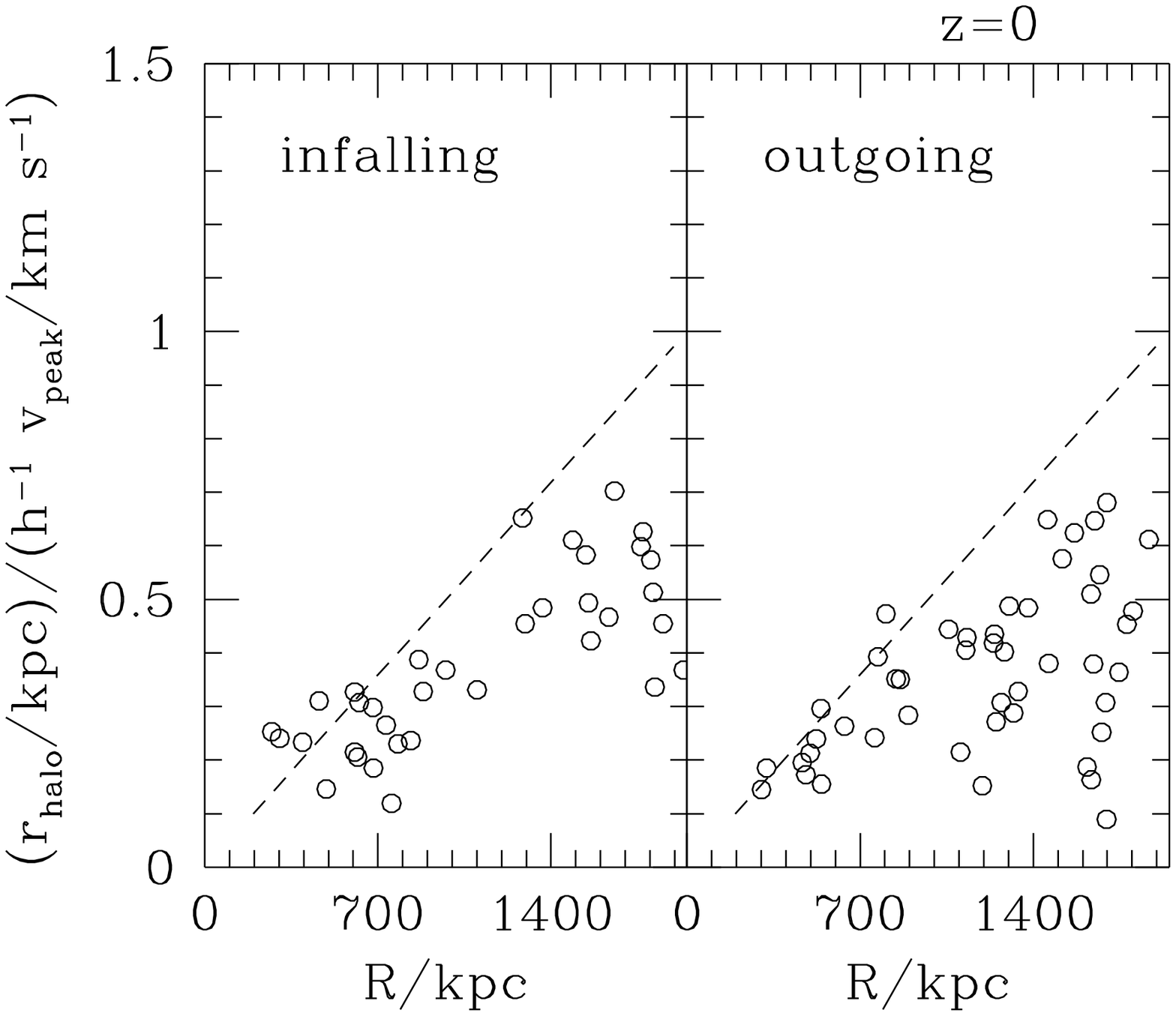}}
\put(0, 70)
{\epsfxsize=8.8truecm \epsfysize=6.truecm 
\epsfbox[40 300 600 720]{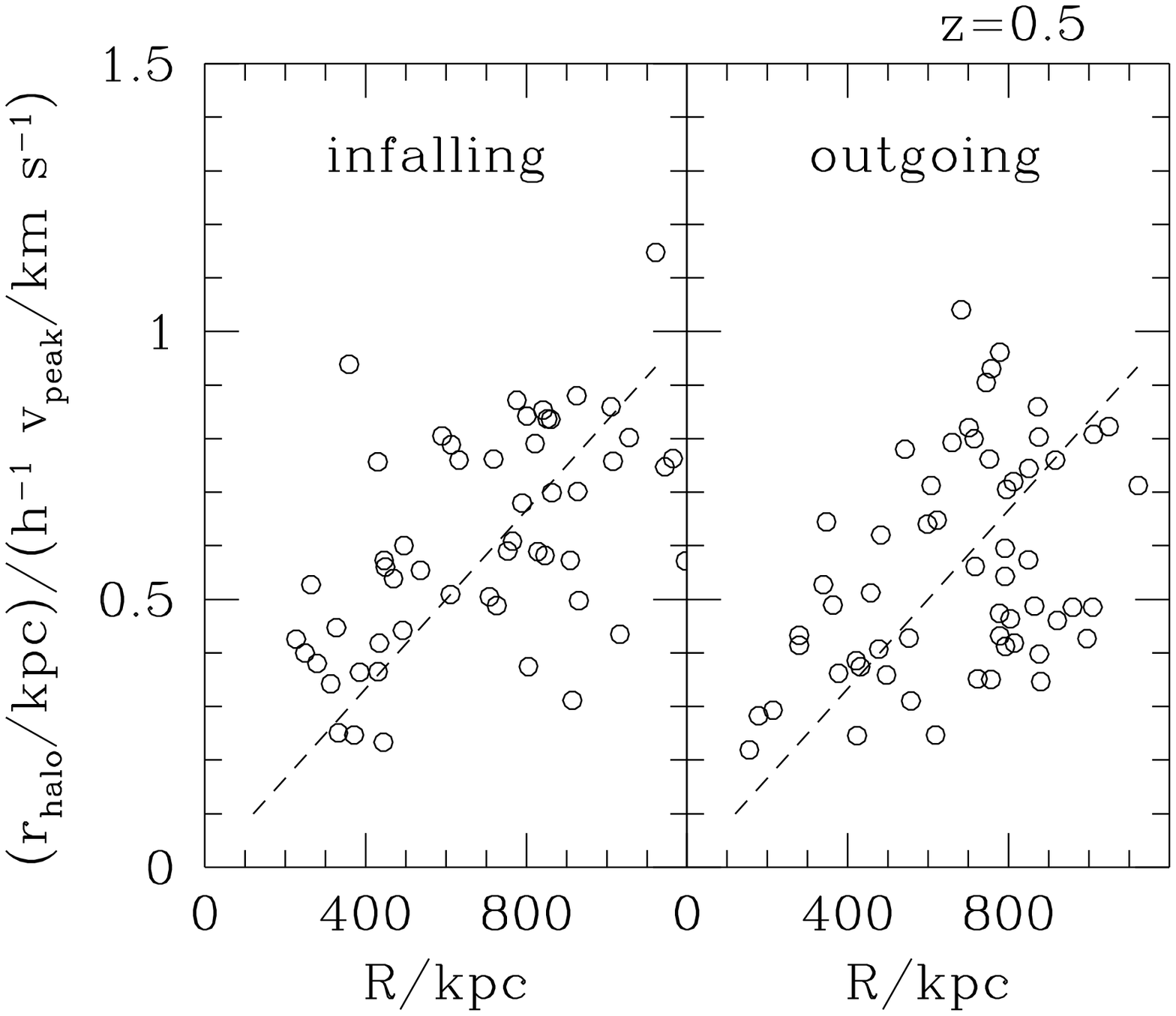}}
\end{picture}
\caption{Same as in Figure~\ref{f:rhalvpk} 
but the halo radii are measured
using the SKID algorithm, that discards unbound particles,
and data are shown for $z=0.5$ too (lower panels). Only 
data for cluster halos are plotted; at $z=0.5$, the formal 
cluster's virial radius is $R_{200}=1200\kpc$.}
\label{f:rhalvpkS}
\end{figure}
The corresponding results using SKID halo radii are shown in
Figure~\ref{f:rhalvpkS} which also shows results for $z=0.5$
(only for halos with $R<R_{200}$; note the 
different scale on the horizontal axis from $z=0$ to $z=0.5$;
$h=h_0(1+z)^{1.5}$ is $H_z$ in units of 100 Mpc$/\kms$).
The results for $z=0$ confirm the general 
picture illustrated by the previous Figure~\ref{f:rhalvpk}, but
here the points have a larger scatter
and the trend with $R$ is weaker, especially for outgoing
halos: this is expected because, in the
``dynamical'' definition used by SKID, the tidal radii are less
sensitive to $\rho_{bkg}(R)$. At $z=0.5$, 
there is  evidence that  tidal effects are already operating  at
this relatively high redshift, 
in agreement with the trend shown by Figure~\ref{f:rhal-R}.

\subsection {Fraction of mass in halos}

\begin{figure}
\centering
\epsfxsize=\hsize\epsffile{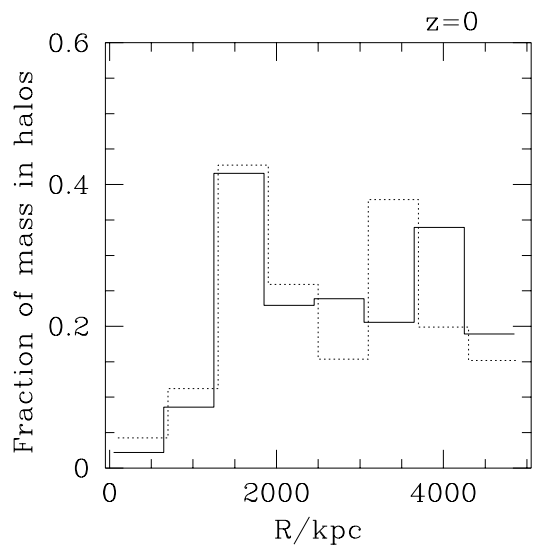}
\caption{Fraction of mass within resolved halos in spherical shells
of thickness 600 kpc
versus cluster-centric position at $z=0$. The solid line is for RUN1 
and the dotted line is for RUN2 (slightly shifted for clarity).}
\label{f:mfrac}
\end{figure}
Figure~\ref{f:mfrac} shows the fraction of the mass attached to halos.
As expected, it decreases sharply approaching the cluster center: it
is $\mincir 5$\% at $R\mincir$ 500 kpc and increases to $\sim 20$\% at
$R_{200}$. The total fraction of mass attached to halos within $R_{200}$
is about 13\%.
(These values do not depend sensitively on the adopted
value of the softening parameter.)  Outside the cluster, the halos
account for about 20\% of the total mass. The peak at $R\sim$ 1.5 Mpc
is not significant: it is caused by the largest halo within the
cluster of mass $2.3\cdot 10^{13} \Msun$ that contributes alone 5\%
of the total mass of the cluster and half of the mass in that bin.

We can compare the mass fraction of peripheral halos with the
Press--Schechter approximation (1974), using again a
routine supplied by P. Tozzi.  Adopting a minimum halo mass of
$1.35\cdot 10^{10} \Msun$ (16 particles) and the mass of the most
massive peripheral halo ($3.4\cdot 10^{12}\Msun$), we find $\sim 0.25$
from the analytical theory, in good agreement with the average of the
last five outer bins, $\sim 0.23$.

\subsection{Merging and `surviving' halos}

Comparing the distribution of the halos at $z=0.5$ and $z=0$, we can
determine the merger rate and the fraction of 
halos that dissolve in this interval.
We selected all the halos (of RUN1) with more than 32 particles at
$z=0.5$ and examined their association with halos with more than 16
particles at $z=0$.
We determine that a high--$z$ halo is the progenitor of a low--$z$
halo if the latter contains at least a significant fraction of the
mass of the former, say 25\%. However, to 
account for mass loss via tidal stripping we need be 
less restrictive, therefore we considered also an extreme mass fraction 
cut--off of 1\%,
but our results are not very sensitive to this limit.

We find, for halos with apocenters $r_{apo}\ge R_{200}$ and at least one
progenitor among those identified at $z=0.5$, that 5--9\% of them,
depending on the mass fraction cut--off, 
are the product of a merger. 
There are a large number of mergers among halos with apocenters
close to $R_{200}$ (3--5 out of 17). 
However, of the 38 halos with $r_{apo}\mincir 80$\%$\,R_{200}$, 
none has merged. 
We find no evidence for merging once the halos have entered the cluster. 
This seems to argue against the possibility of Butcher--Oemler (1978, 1984) 
galaxies being cluster members with star formation switched on by mergers
in the cluster environment (Couch \etal 1994).

\begin{figure}
\centering
\epsfxsize=\hsize\epsffile{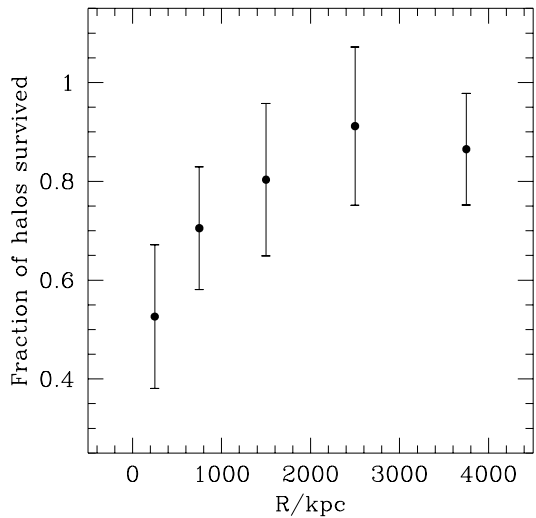}
\caption{Fractions of z=0.5 halos above resolution that have survived to
$z=0$  retaining more than $\sim 50$\% of their
original masses; the errorbars are 1--$\sigma$ Poissonian uncertainties
estimated from the numbers of halos in each distance interval.}
\label{f:SurviveR.01.z05}
\end{figure}

Figure~\ref{f:SurviveR.01.z05} 
shows the fraction of halos identified at $z=0.5$ 
in different distance intervals
that are also associated with $z=0$ halos. Although we adopt the extreme
 1\% mass fraction cut--off to establish this association, the 
results shown in the Figure essentially give the probability that 
halos will survive with 50\% of their original masses attached. 
In fact, $z=0.5$ halos with masses just above the limit of 32
particles will be resolved at the 16--particle limit of the
$z=0$ sample, only if they have not lost more than 50\% of their
masses; and these small halos dominate the statistics.
At $R\mincir 0.5 \Mpc$, such probability is $\sim 0.5$, against 
a value of order unity in the cluster periphery. 

\subsection{Orbital parameters}
\label{ss:orbits}

We now shift our attention to the motions of the halos within the cluster.
Halos that follow radial orbits are more likely to be disrupted than
halos that have circular motions, since the former penetrate further
into the cluster potential well.  Do we
detect a bias in favour of circular orbits?  We calculate orbits
by approximating the cluster potential as a spherical static potential,
computed using the density profile at $z=0$ and then using the position
and velocity information for the halos.
For comparison, we also compute the orbits of a random subset 
of 20,000 particles.

\begin{figure}
\begin{picture}(300, 430)  
\put(0, 280)
 {\epsfxsize=8.truecm \epsfysize=6.truecm 
\epsfbox[40 300 600 720]{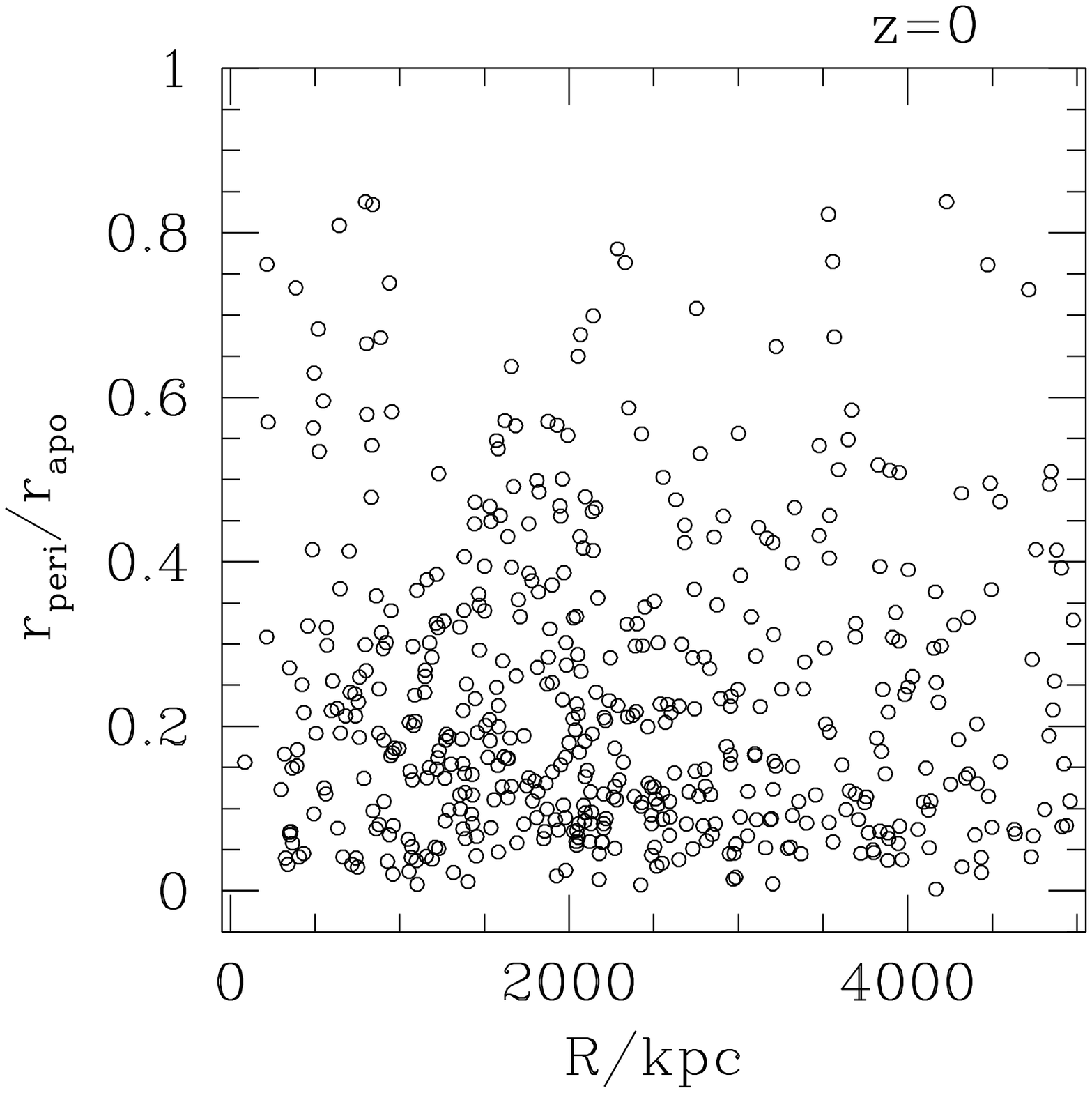}}
\put(0, 70)
{\epsfxsize=8.truecm \epsfysize=6.truecm 
\epsfbox[40 300 600 720]{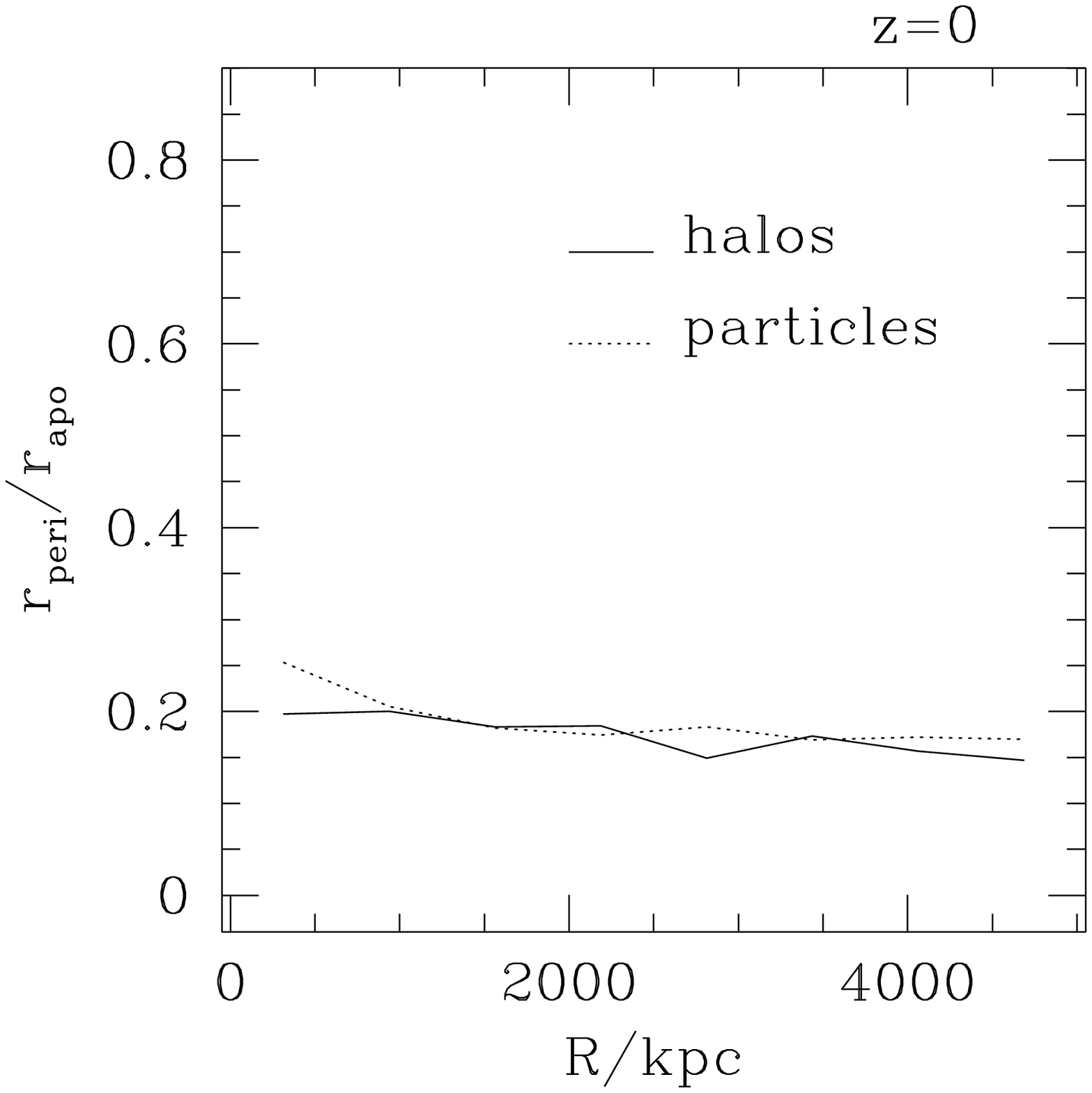}}
\end{picture}
\caption{The upper panel shows the ratio pericentric and apocentric
distances for each halo plotted against its current position.
The lower panel shows the average value of these ratios for the halos
(dashed curve) and a random subsample of cluster dark matter particles
(solid curve). The errobars are the dispersion about the average.}
\label{f:peri-apo}
\end{figure}
In the upper panel of Figure~\ref{f:peri-apo} we plot the ratios
$r_{peri}/r_{apo}$ vs $R$ for the halos of RUN2 (RUN1 gives identical
results). The lower panel shows the average values of this ratio for
the halos (solid line) and the particle sample (dotted line). The
errorbars show 1--$\sigma$ standard deviations. We find that radial
orbits are quite common and circular orbits are rare. The median ratio
of apocenter:pericenter is approximately 6:1 
and nearly 25\% of the halos are
on orbits more radial than 10:1.  
A rough calculation by the authors reveals that this is very close 
to an isotropic orbital distribution within an infinite isothermal 
potential.

We do not detect a large difference between halo orbits as a function
of $R$, nor do we find a difference between the orbits of the particle
background and the halos. This is surprising since we expected to find
fewer halos on radial orbits near the cluster center. The expected
bias could be disguised if the central overmerging problem originated
within the dense clumps that formed before the final cluster. Finally,
we note that the radial velocity dispersion of the halos within the
cluster is $720 \kms$, a value that is within a few percent of the
dispersion of the background particles \--- even when only the most
massive halos are considered.

\begin{figure}
\centering
\epsfxsize=\hsize\epsffile{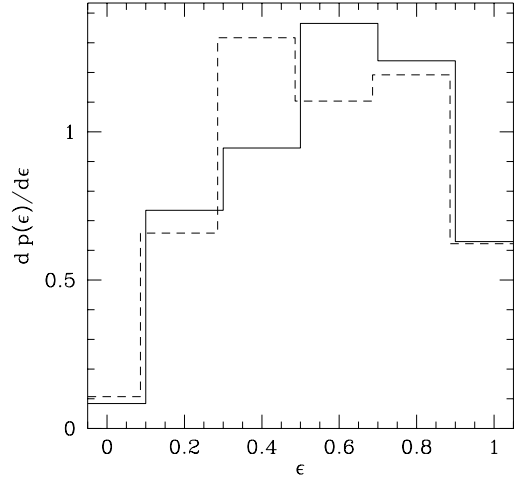}
\caption{
Probability density distribution of halo ``circularities''
$\epsilon\equiv J/J_C(E)$ (see text) for {\sl cluster}
($R<R_{200}$; solid line) and {\sl peripheral} halos (dashed).  
}
\label{f:ecc}
\end{figure}
\begin{figure}
\centering
\epsfxsize=\hsize\epsffile{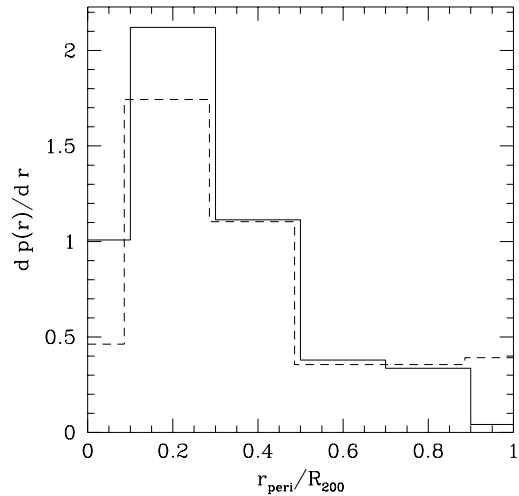}
\caption{
Probability density distribution of halo pericenters $r_{peri}$
(line styles as in Fig.~\ref{f:ecc}).
}
\label{f:peri}
\end{figure}
Information on the distribution of the orbital parameters of halos 
can be used to model the effects of tidal stripping and
dynamical friction for halos within halos in 
semi--analytic models of structure formation based on the
Press--Schechter approximation. 
Figure~\ref{f:ecc} and \ref{f:peri} plot the probability density distributions
of ``circularities'' and pericenters for cluster (solid line) and 
peripheral (dashed) halos. For each
halo, the ``circularity'' $\epsilon\equiv J/J_C(E)$ is defined as the ratio
of its angular momentum to that of a circular orbit with the same
energy (Lacey \& Cole 1993). There are no marked differences between
the two groups of halos, although the orbits of cluster halos
are more close to circular and penetrate further into the cluster than
those of peripheral halos. Among the latters, 15\% have 
pericenters outside the cluster's boundaries and 9\% come as close
as 200 kpc ($0.1 R_{200}$) to the cluster's center. This condition is
twice more frequent among cluster halos; in the whole sample the
fraction of $r_{peri}< 200\kpc$ is 14\%. These results are in good agreement
with results presented by Tormen (1997).

\begin{figure}
\begin{picture}(300, 430)  
\put(0, 280)
 {\epsfxsize=8.truecm \epsfysize=6.truecm 
\epsfbox[40 300 600 720]{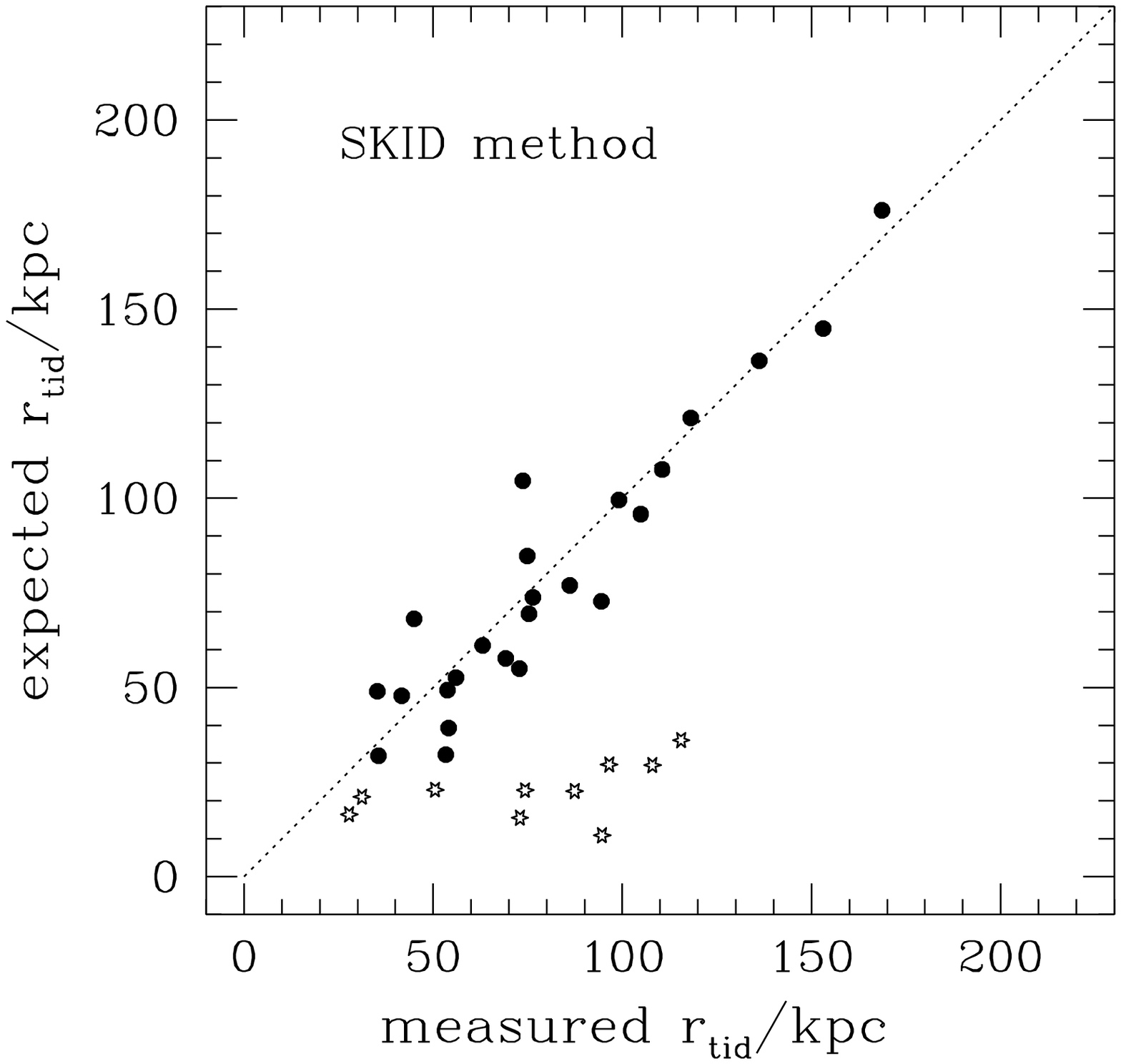}}
\put(0, 70)
{\epsfxsize=8.truecm \epsfysize=6.truecm 
\epsfbox[40 300 600 720]{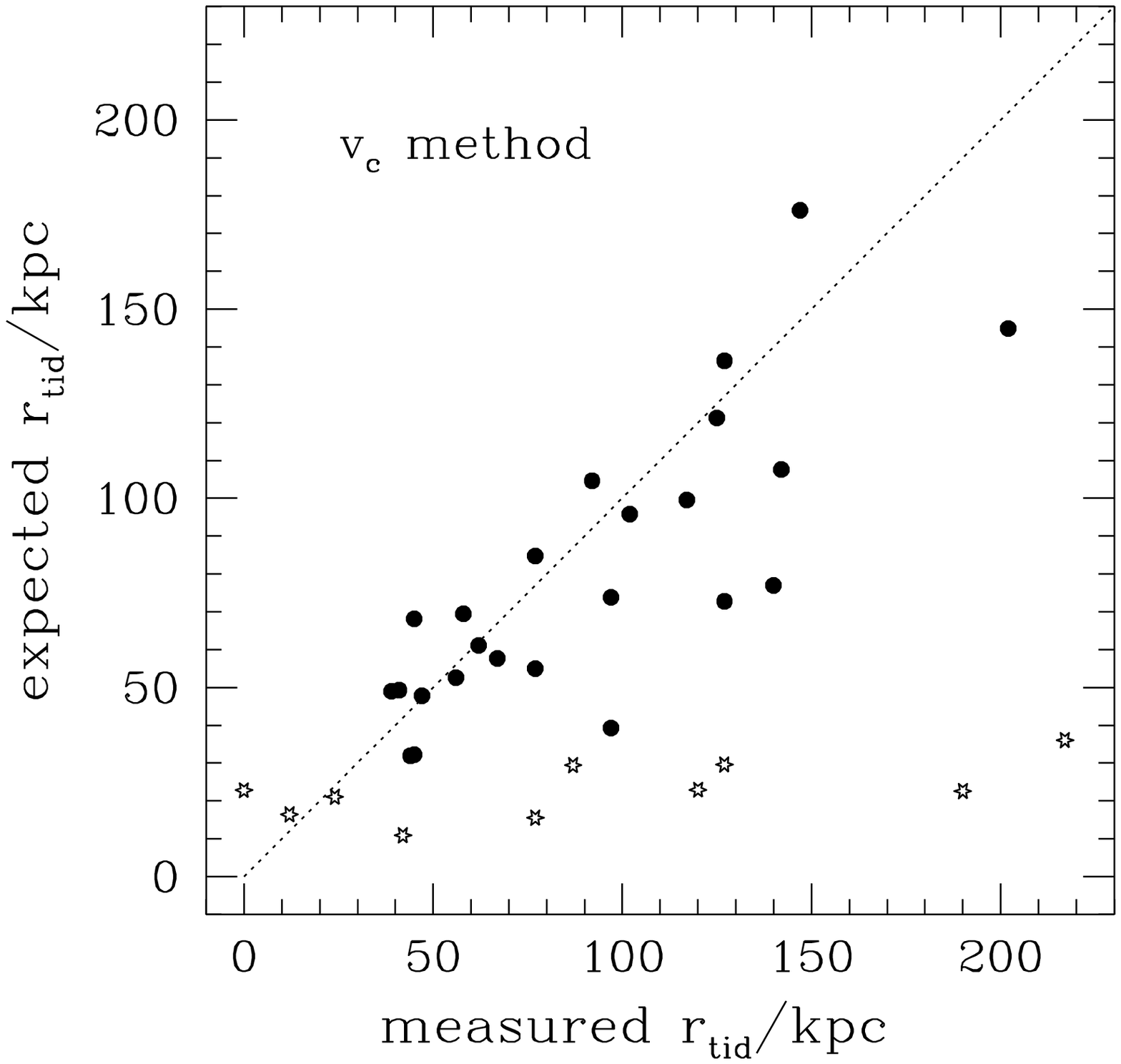}}
\end{picture}
\caption{We plot the measured values of halo tidal radii against their 
expected values, assuming that the halos have isothermal mass
distributions that are tidally stripped at their pericentric
positions.  The upper panel uses halo tidal radii measured using SKID,
therefore only self--bound particles are included to determine halo
sizes; the `$v_c$' method uses halo circular velocity profiles and
includes the cluster background. The points represent outgoing halos
(at $z=0$) with $R\le 0.8 R_{200}$; the stars denote those with
$r_{peri}<300\kpc$.}
\label{f:rexp}
\end{figure}
For such radial orbits, 
we expect that the tidal radii of the halos are determined primarily
by the global tidal field of the cluster. We can check if this is
correct by estimating the truncation radius at the pericenter of each
halo using $r_{tid}\simeq r_{peri}\cdot v_{peak}/V_c$ (the use of
$V_c(R)$ at $R=r_{peri}$, instead of the constant value
$V_c=V_{200}=1000\kms$ does not make any detectable difference,
because the variation of $V_c(R)$ is $\mincir 10$\% in the range $0.05
< R/R_{200}<1$ ). We can test this prediction for our outgoing halos
that must have passed pericenter
recently, enhancing the likely validity of our approximation. 
In Figure~\ref{f:rexp} we plot
the expected tidal radius, according to the above formula, against the
value measured, for both methods (from SKID and from the circular
velocity profile $v_c(r)$). The agreement is excellent for the SKID
values, with the exception of the points marked as stars.  These
latter points represent halos that are on very eccentric orbits such
that $r_{peri}$ is less than $300\kpc$: These halos are more likely to
suffer impulsive collisions with other halos as they pass close
to the cluster center.
We note that many of these halos have tidal tails 
that may cause the measured tidal radii to be overestimated. 
The scatter in the correlation increases
when we measure halo sizes using the halo circular velocity
profiles, but the trend is still apparent.
The radii estimated with this method are more sensitive to
tidal tails than SKID radii, since the latter can discard the unbound
streams of particles. This is the reason why some points 
in the lower panel of the Figure correspond to larger $r_{tid}$ than
those in the upper one.

\begin{figure}
\centering
\epsfxsize=\hsize\epsffile{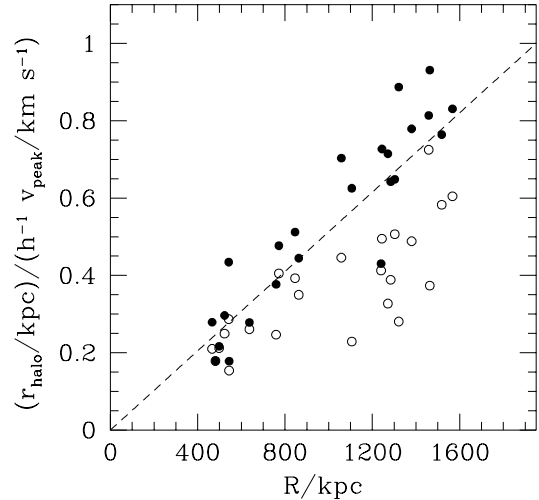}
\caption{The dashed line gives the expected dependence on $R$ of the ratios
$\eta\equiv r_{halo}/v_{peak}$ for isothermal halos that are tidally
limited by the cluster at a distance $R$ as they fall into the cluster
for the first time.  Such ratios attain a minimum when halos reach
pericenter and are maximally stripped; therefore, moving away from the
cluster's center, outgoing halos have values of 
$\eta$ systematically below the
line (the open circles; same as in the upper--left panel of
Figure~\ref{f:rhalvpkS}). If we `correct' their $\eta$s using the
information on their pericenters, the points are brought back to the
line (the filled circles).}
\label{f:rhalvpkScor}
\end{figure}
Recalling Figures~\ref{f:rhalvpk} and 
\ref{f:rhalvpkS}, we can now check if tidal
stripping is indeed responsible for the low values of $\eta\equiv r_{halo}/
(h^{-1} v_{peak})$. 
Using the information on $r_{peri}$, we can `correct'
the values of $\eta$ 
for outgoing halos
as $\eta\rightarrow \eta\cdot R/r_{peri}$.  The
effect of this correction is shown in Figure~\ref{f:rhalvpkScor}, in
which the open circles are the original points of the upper--right
panel of Figure~\ref{f:rhalvpkS} and the filled circles are `corrected'
As expected, the corrected $\eta$ scale as $R/R_{200}$.
In conclusion, the isothermal model predictions for the tidal radii
work well when the pericentric positions of the halos are
used.  In turn, this good agreement confirms that our estimates of the
orbital parameters for cluster halos are correct.

\section{Halo internal properties}
\label{s:pint}

We shall now examine the internal structure of halos in more detail 
with the main objective of studying how the cluster environment affects them.
 In particular, we shall focus on the distribution of
$r_{peak}$ and $v_{peak}$, that provide information on their internal
concentrations.
Then, we shall
study in detail
a small subsample of well
resolved halos
and examine
their density profiles and the compatibility with the analytic fits
generally adopted in the literature to describe dark matter halos 
(Hernquist 1990; Navarro, Frenk \& White 1996).

\subsection{Statistical distribution of halo properties}

A detailed analysis of isolated cold dark matter halos in $N$--body
simulations has been carried out by Navarro, Frenk \& White (1996 and
1997, NFW in the following).  Their simulations had a mass resolution
such that individual halos contained of order 10,000 particles and
force softening that was 1\% of the final virial radii.  Over a wide
range of masses, NFW found relations between the properties of their
profiles, $r_{peak}$ and $v_{peak}$ and their virial masses
$M_v$. Furthermore, the density profiles of halos could be fit by a
universal formula with varying concentrations, $c$, that can be
predicted from their masses within a given cosmological model.

Here we want to address two questions:

$(i)$ Do the same relations found by NFW for
isolated halos hold for the {\sl peripheral halos}
that surround
the virialized cluster?  These 
are relatively isolated, but they evolved in an 
environment that is special owing to the nearby cluster.  
For instance, the
streaming motions it induces could have determined peculiar merging
histories of the peripheral halos, and merging may play a crucial role
in shaping their internal structures (cfr.  Syer \& White 1997; Moore
\etal 1997).

$(ii)$ Do the same relations apply for {\sl cluster halos}, that are
affected by tidal stripping and halo--halo encounters?
Cluster halos could also have peculiar properties reflecting the 
high--density environment in which they formed.

These two factors could then affect the properties of 
our halo sample 
in such a way that, for instance, the ratios $v_{pk}/r_{pk}$, that reflect
the concentrations of the halos, are systematically higher than for
the field population (e.g. cluster halos form earlier or
are biased towards high concentrations such that they have survived
tidal disruption)
Alternatively, impulsive mass loss may cause halos to subsequently ``relax''
and re--expand 
towards an `equilibrium' configuration. This could be detected as a
change in concentration versus orbital position.

We shall focus on the following two relations, for cluster and peripheral
halos, $M_{halo}$ vs. $v_{peak}$ and $r_{peak}$ vs. $v_{peak}$. The former has
the interest of relating masses to velocities, that are in
principle more easily observable, whilst the latter gives information
on halo concentrations and the related issues discussed above.

\begin{figure}
\begin{picture}(300, 430)  
\put(0, 280)
 {\epsfxsize=8.truecm \epsfysize=6.truecm 
\epsfbox[40 300 600 720]{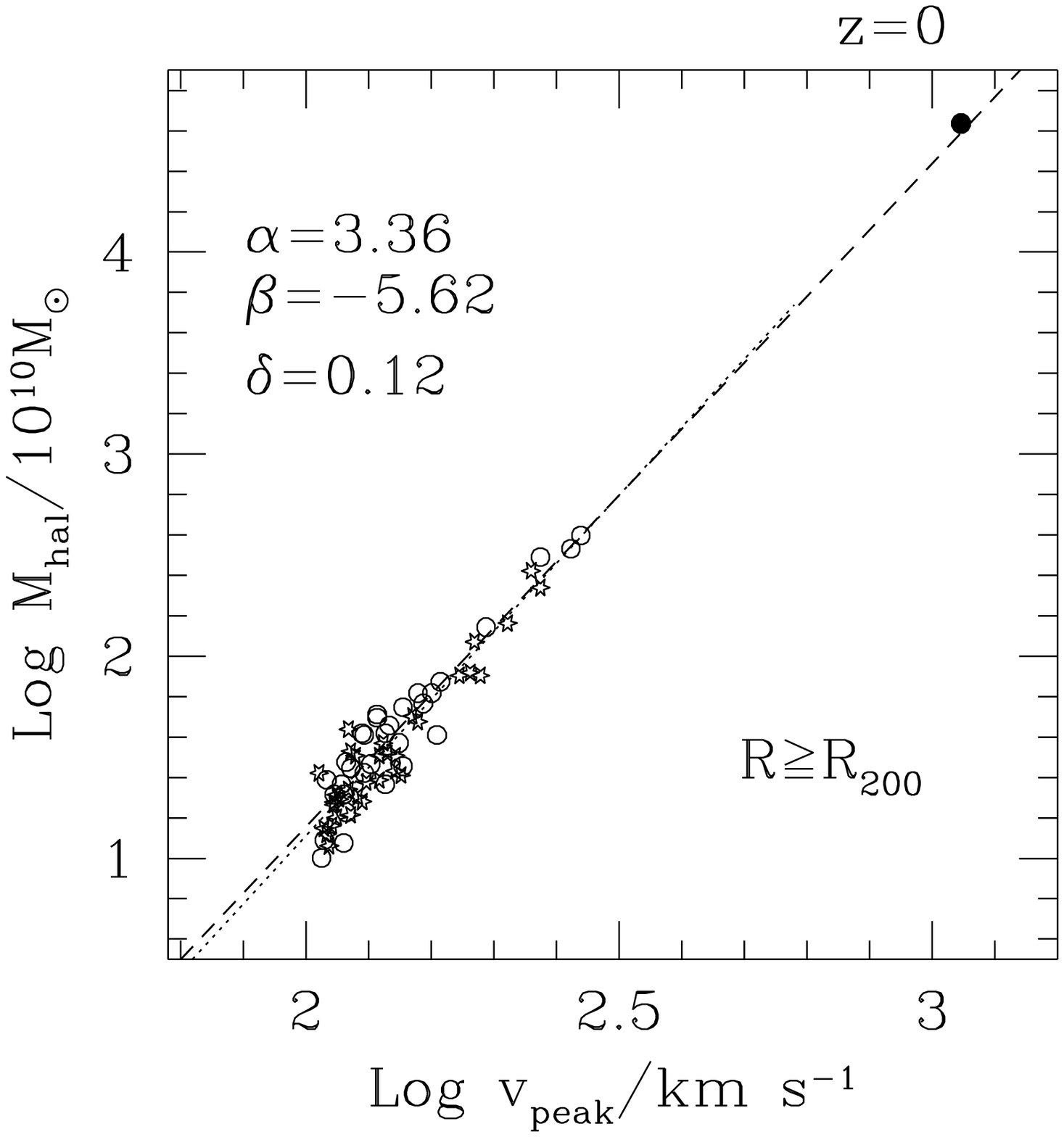}}
\put(0, 70)
{\epsfxsize=8.truecm \epsfysize=6.truecm 
\epsfbox[40 300 600 720]{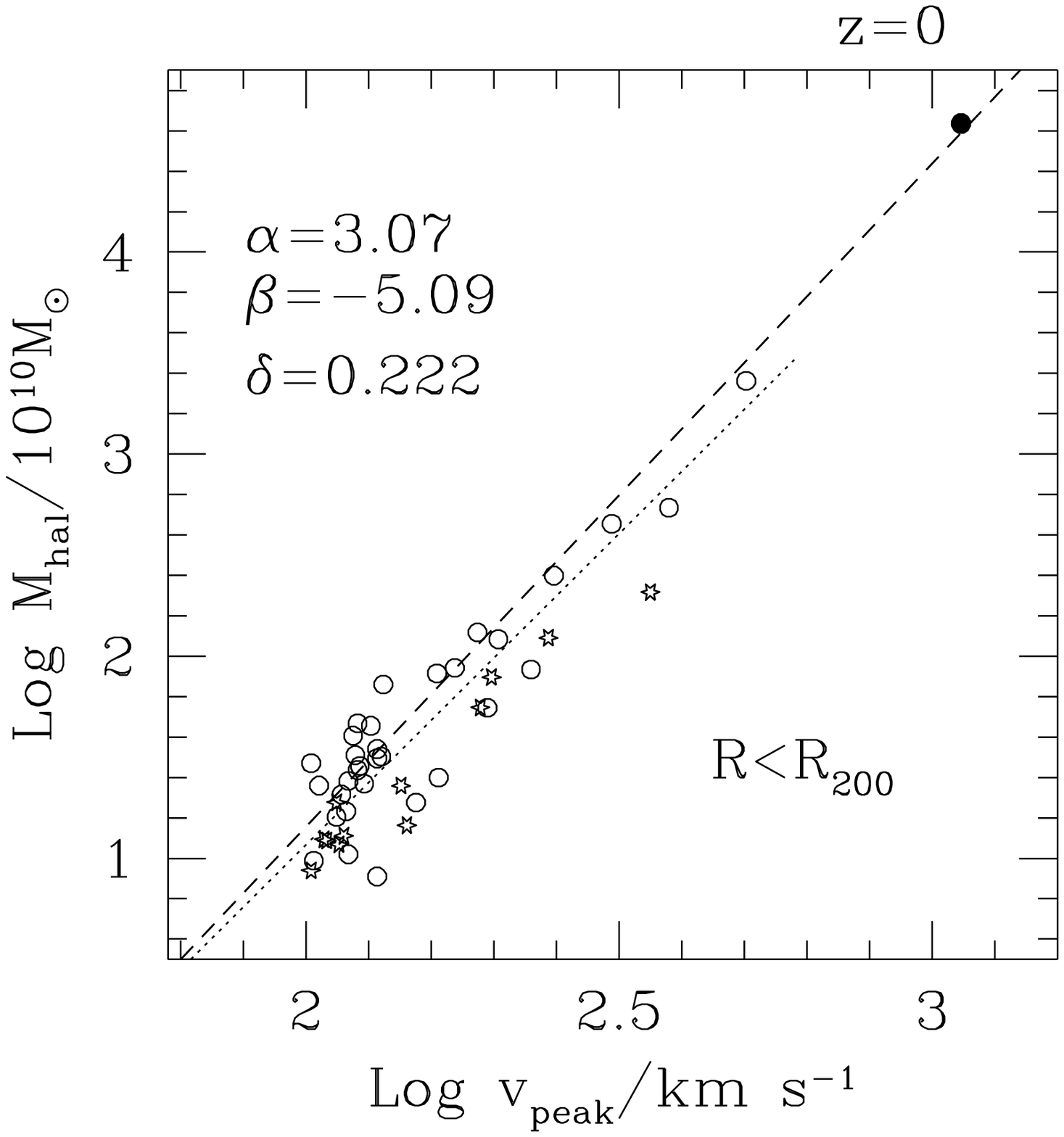}}
\end{picture}
\caption{
Distribution of halo masses versus $v_{peak}$ for `peripheral'
(upper panel) and cluster halos (lower panel) at $z=0$. 
The stars mark inner points,
with $R<1.5R_{200}$ and $R<0.6R_{200}$, respectively for the two groups.
The filled circle gives the values for the entire cluster. 
The dashed line is the power--law fit obtained by Navarro \etal (1996) for
isolated halos.  
The dashed lines are similar power--law fits to the data
of the form $M=10^\beta v^\alpha$, where $\delta$ is 
the rms scatter in mass about the fit.
}
\label{f:Mass-vpk}
\end{figure}
The distribution of halo masses vs. $v_{peak}$ is illustrated in
Figure~\ref{f:Mass-vpk}, for peripheral (upper) and cluster halos 
(lower panel)
and compared with the relation found by NFW within the cosmological
model we have adopted (NFW 1996; dashed line).
(For clarity, only points with $v_{peak}>100\kms$ are shown, since the
noise is large for small halos).
In each panel, the stars mark {\sl inner} halos
of each group 
(see the caption of the Figure for details; 
note that some inner halos within the present ``periphery'' sample 
have highly radial
orbits and were inside the cluster at an earlier epoch.)
For peripheral halos, the agreement between our data and the NFW
curve is excellent
(note that most of the scatter owes to points with
$v_{peak}<150\kms$ that may be 
affected by the numerical noise).
Cluster halos have a much larger scatter and the fit is shallower.
Their masses are smaller for a given $v_{peak}$, as
naturally expected for stripped halos and this is most noticeable
for the {\sl inner} cluster halos.

\begin{figure}
\begin{picture}(300, 430)  
\put(0, 280)
 {\epsfxsize=8.truecm \epsfysize=6.truecm 
\epsfbox[40 300 600 720]{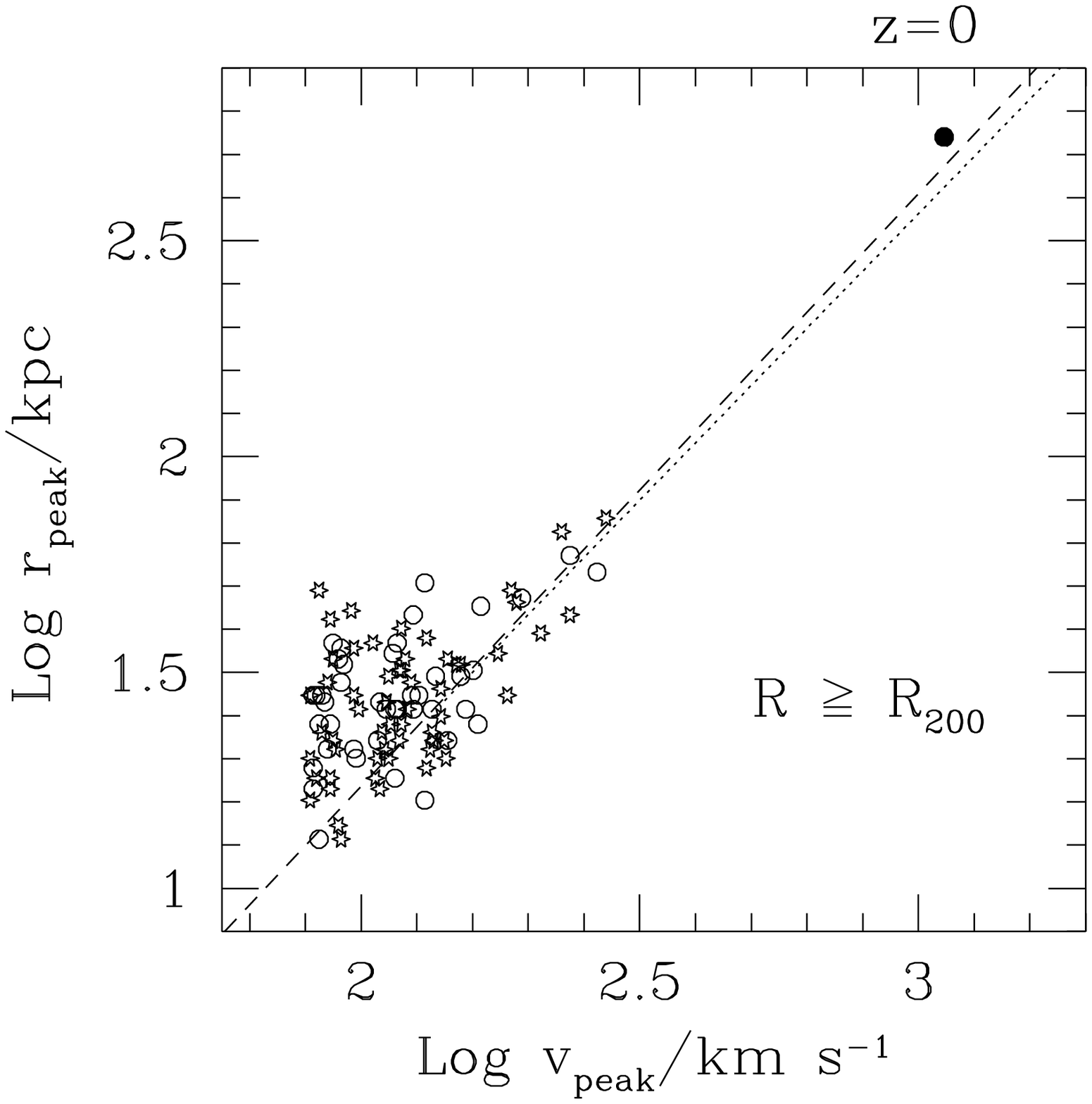}}
\put(0, 70)
{\epsfxsize=8.truecm \epsfysize=6.truecm 
\epsfbox[40 300 600 720]{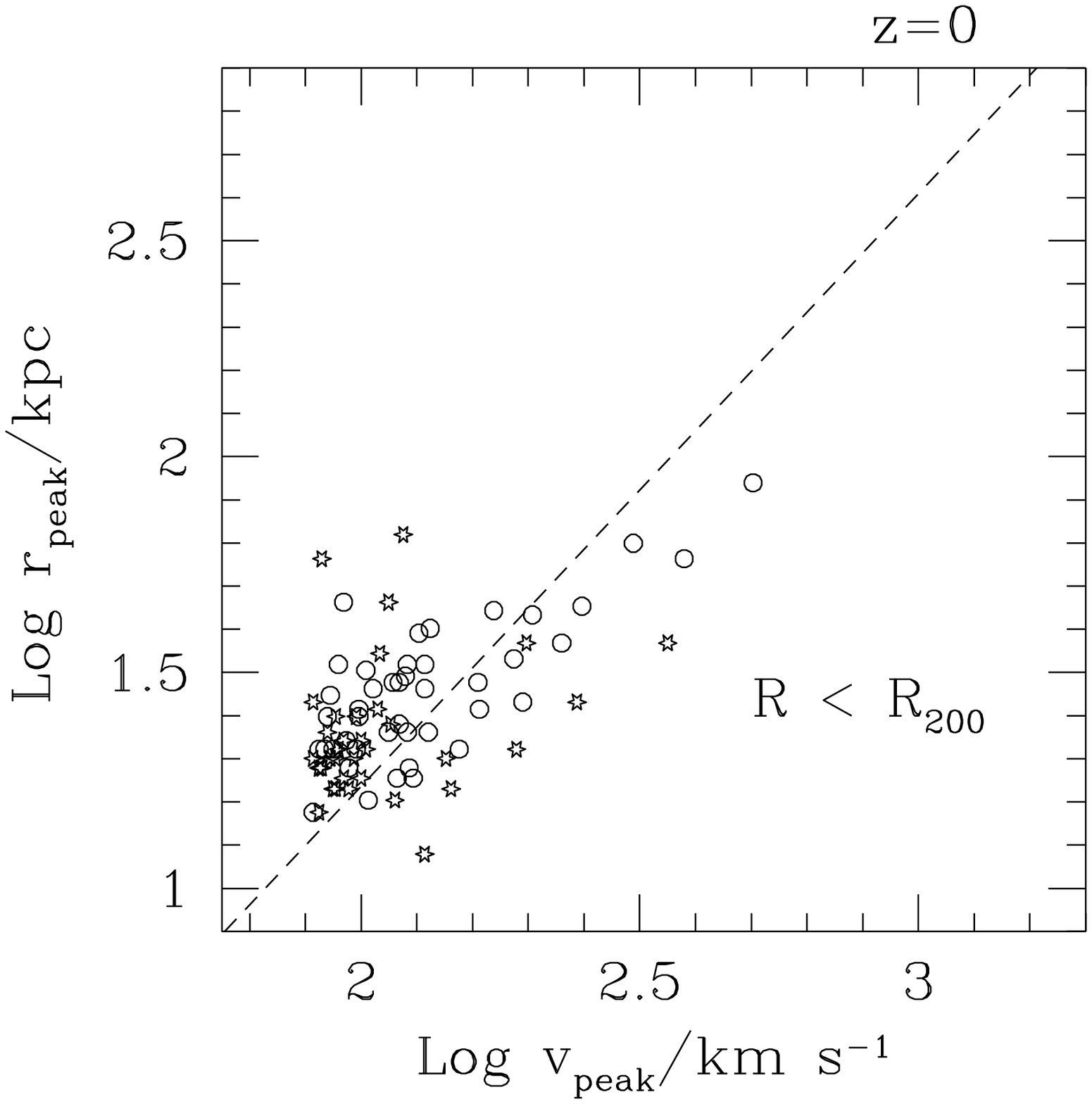}}
\end{picture}
\caption{Distribution of $v_{peak}$ and $r_{peak}$ for peripheral
(upper panel) and cluster halos (lower panel) at $z=0$. The points have the
same meanings as in Figure~\ref{f:Mass-vpk}. 
As before, the dashed line gives 
the expected relation for isolated halos (Navarro \etal
1996) and the dotted line is a fit to the points with $v_{peak}\ge 150\kms$.
}
\label{f:rpk-vpk}
\end{figure}
The distribution of $v_{peak}$ and $r_{peak}$ is shown in
Figure~\ref{f:rpk-vpk}, again for peripheral halos in the upper panel 
and cluster halos in the lower panel.
The dashed line gives the relation for the halos studied by
NFW and the dotted line in the upper panel 
is a power--law fit to the halos with 
$v_{peak}>150\kms$ (that appear to be less affected by noise).
The behaviour of the peripheral halos clearly agrees well, within the scatter,
with that expected from the analysis of NFW. On the contrary, cluster
halos lie significantly below the dashed line: they are skewed towards 
smaller values of $r_{peak}$ for a given $v_{peak}$, with a more
prominent effect for the innermost halos.

\begin{figure}
\begin{picture}(300, 430)  
\put(0, 280)
 {\epsfxsize=8.truecm \epsfysize=6.truecm 
\epsfbox[40 300 600 720]{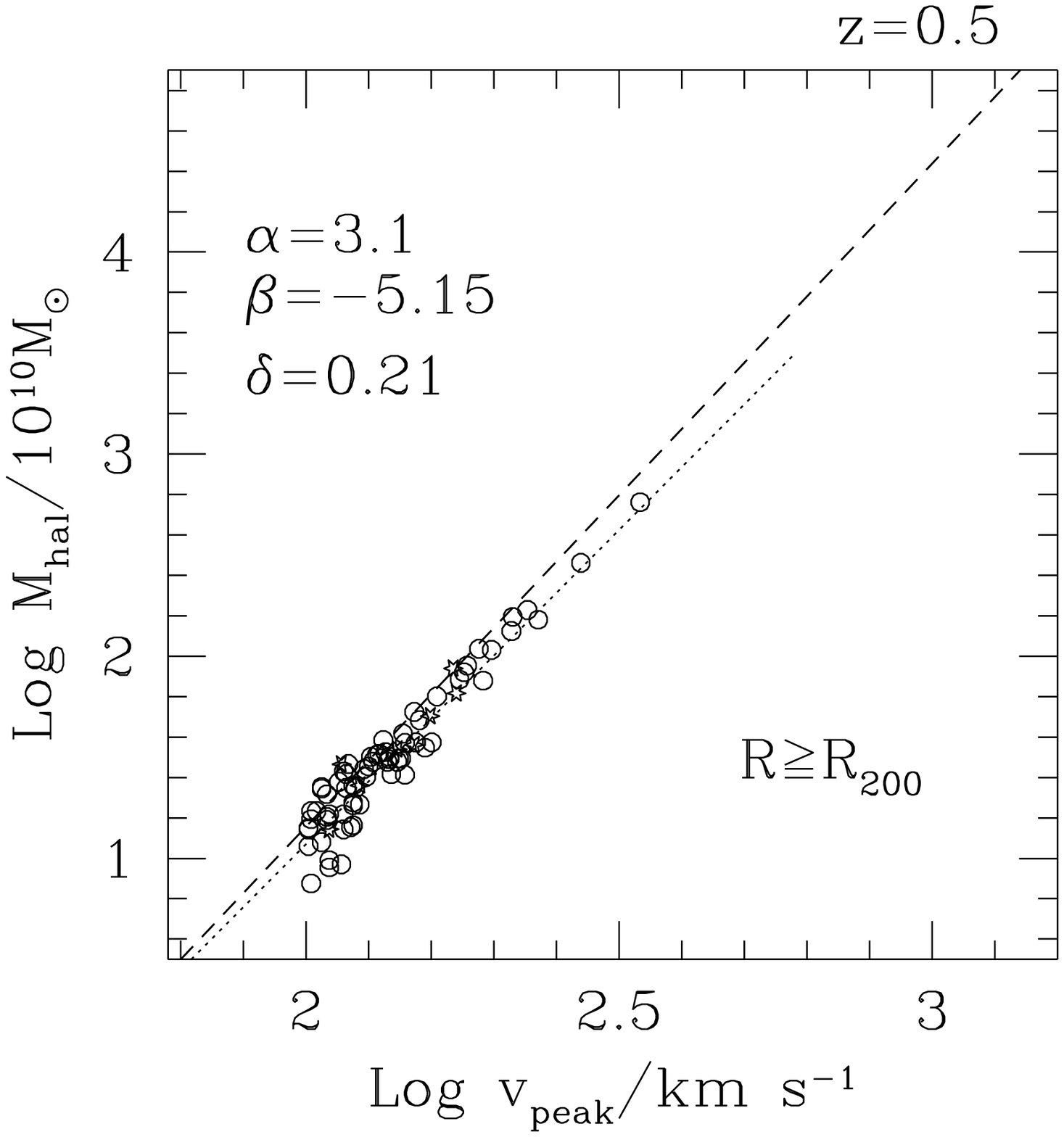}}
\put(0, 70)
{\epsfxsize=8.truecm \epsfysize=6.truecm 
\epsfbox[40 300 600 720]{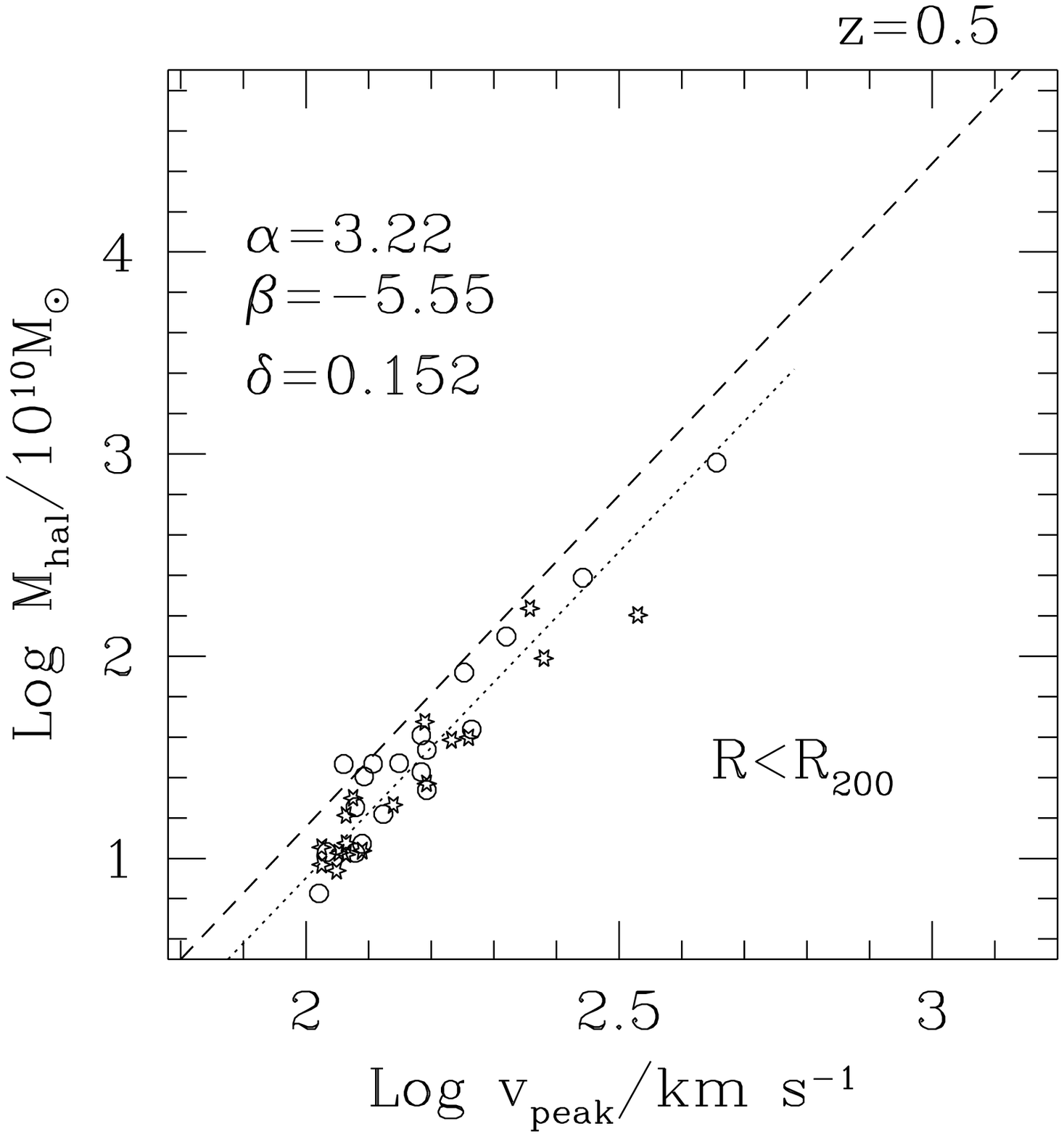}}
\end{picture}
\caption{
Symbols and lines are the same as in Figure~\ref{f:Mass-vpk} but for 
$z=0.5$. Here, $R_{200}$ is the formal cluster's virial radius at that epoch.
}
\label{f:Mass-vpk.z05}
\end{figure}
\begin{figure}
\begin{picture}(300, 430)  
\put(0, 280)
 {\epsfxsize=8.truecm \epsfysize=6.truecm 
\epsfbox[40 300 600 720]{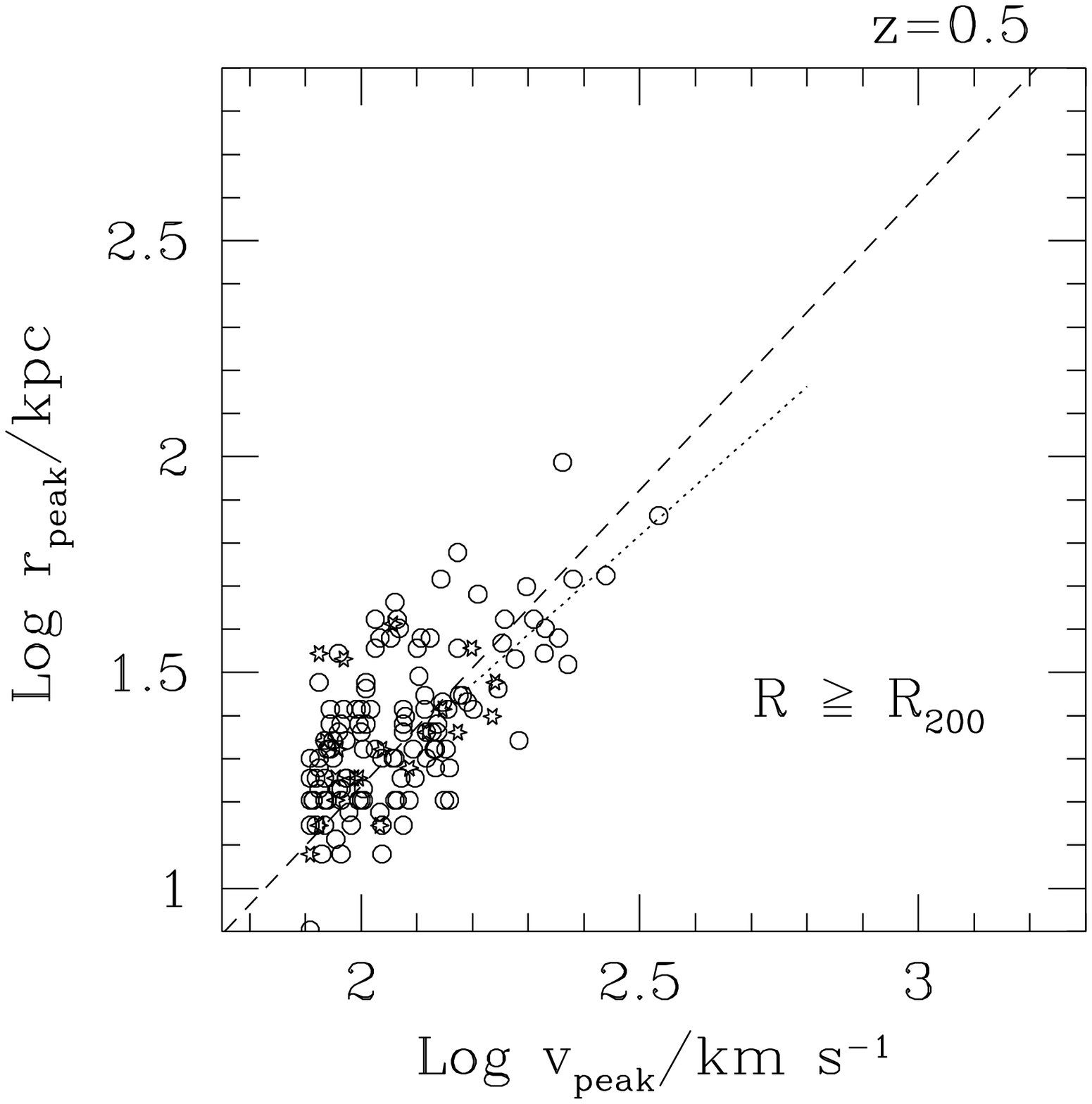}}
\put(0, 70)
{\epsfxsize=8.truecm \epsfysize=6.truecm 
\epsfbox[40 300 600 720]{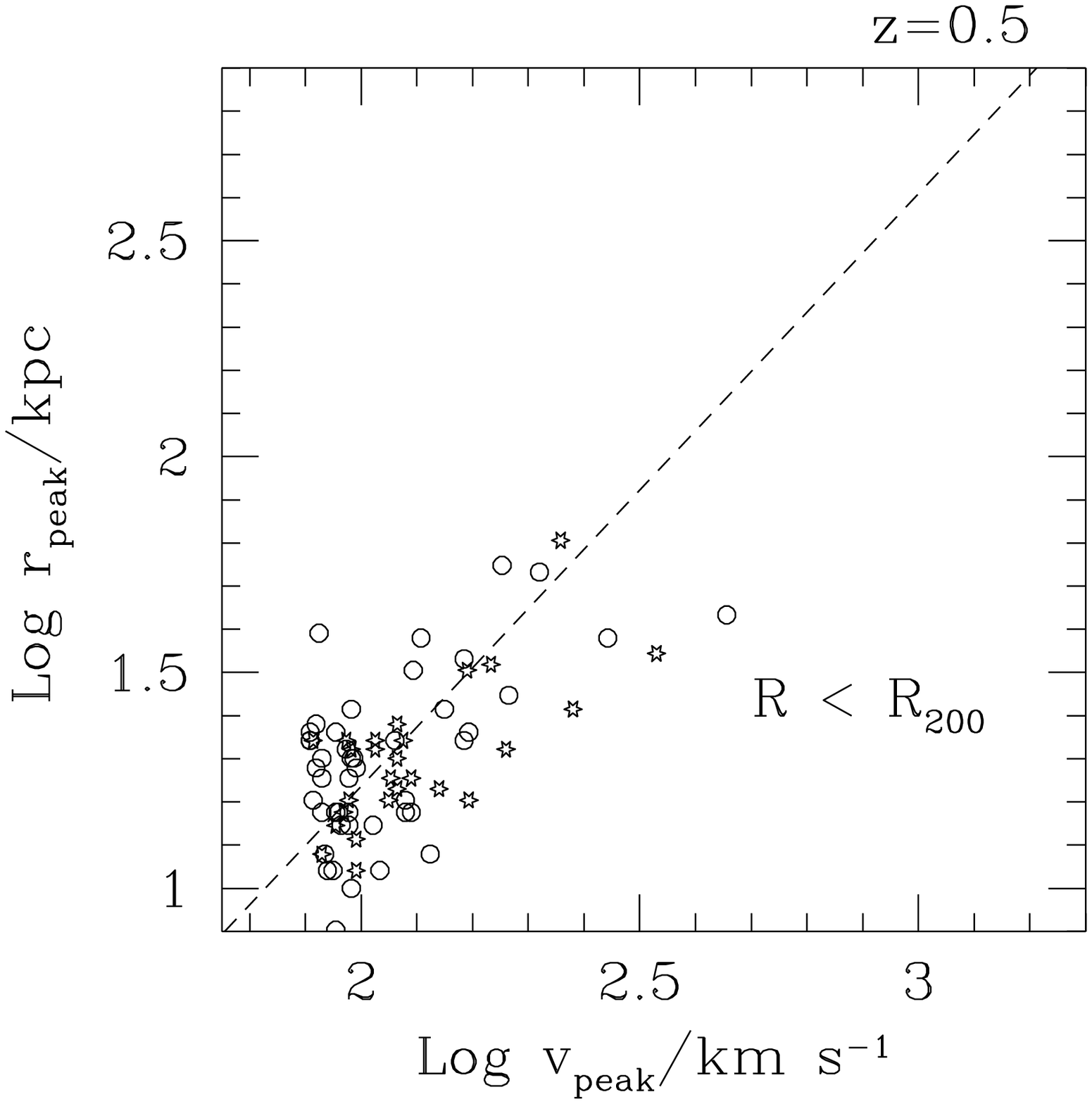}}
\end{picture}
\caption{
Symbols and lines are the same as in Figure~\ref{f:rpk-vpk} but for 
$z=0.5$. Here $R_{200}=1.2\Mpc$ is the formal cluster's virial radius at that epoch.
}
\label{f:rpk-vpk.z05}
\end{figure}
At $z=0.5$, we observe a behaviour similar to that at $z=0$,
as is shown in Figures~\ref{f:Mass-vpk.z05} and~\ref{f:rpk-vpk.z05}
(in these figures, the dotted line is again a fit to the points with
$v_{peak}>150\kms$ and the dashed line, the NFW's relation at $z=0$,
is drawn for comparison).

In conclusion, 
the high--density regions of the
cluster and the ``accelerated collapse'' of the cluster halos
can affect not only the sizes of the halos but also their
internal structures.
We examine this issue further in the next section by looking in detail
at the properties of a sample of large well resolved halos.

\subsection{The large halo sample}

We now restrict our analysis
to several well resolved, massive halos that are 
not as affected 
by force softening or mass resolution, by following the evolution of
those halos that have $r_{peak}>25\kpc$ and $v_{peak}>150\kms$
(from Figure~\ref{f:rpk-vpk}, we can see that these values delimit
a region in the plane $(v_{peak},r_{peak})$ where the properties of
halos behave quite regularly).

To highlight differences between ``cluster halos'' and
``peripheral halos'', we selected two groups in the distance
ranges $R/R_{200}\le2/3$ and $4/3<R/R_{200}<7/3$. Again, these values
of $R$ correspond to those roughly separating different halo
behaviours in Figure~\ref{f:rpk-vpk}. These selection criteria yield 7
halos in the first distance range 
(although names like ``Yogi'', etc$\dots$ would be preferable, for
brevity's sake, we shall simply call them: A, B, C, D, E,
F) and 7 in the second one (G, H, I, L, M, N, O).
The first 3 of each group lie in the inner part of their distance
ranges ($R \mincir 0.8\,$Mpc and $2.8\Mpc\mincir R\mincir 3.5\Mpc$,
respectively).  All of these halos contain at least 1000 particles
within $r_{halo}$.
We excluded halo ``O'' because it is the product of a recent merger
($z\sim 0.5$). 
All of the other halos have only one progenitor at $z=0.5$ and the
inner halos are well defined up to $z\sim 2$.

\subsubsection{Evolution of the profiles}

We now examine the evolution of the density profiles for this
sample of halos. We consider four values of the redshift: 
$z=0$, $z=0.5$, $z=1$, and $z=1.8$.  
In our sample, all peripheral and {\sl outer} cluster halos
show evidence of major mergers at $z\sim 1$, but the inner cluster
halos (A to D) have a well defined major progenitor up to $z=1.8$
(although they have captured many small satellites and experienced
minor mergers since $z\sim 1$).  For peripheral and outer cluster halos we
limit our analysis to $z=0.5$, while for the innermost ones
we will examine the profile data up to $z=1.8$.

\begin{figure}
\centering
\epsfxsize=\hsize\epsffile{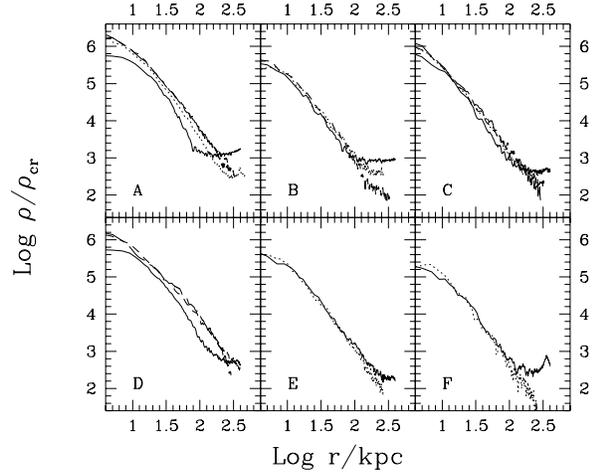}
\caption{
Evolution of the density profiles of a sample of massive 
cluster halos after their last major mergers.
The solid, dotted, short--dashed and long--dashed lines
are the profiles measured at $z=0, 0.5, 1, 1.8$ respectively
(some halos have their last 
major mergers at $z\ge 1$ and correspondingly, not all lines are shown;
the $z=0.5$ profile of D could not be measured because this halo
was too close to the cluster's center at that epoch).
The radius $r$ is the distance from the halo centers in physical kpc and
$\rho$ is measured in units of today's critical density $\rho_{0,cr}$.}
\label{f:rhoevol.1}
\end{figure}

Figure~\ref{f:rhoevol.1} shows the evolution of 
$\rho(r)$ for the cluster halos.
All the curves for the present epoch flatten at large $r$ where the smooth 
particle background density of the cluster, $\rho_{bkg}$, starts dominating
(halos from left to right in each row of figures have increasing
distances from the cluster center).  Such flattening also appears in
the profiles of the $z=0.5$ progenitors of A, B and C because they are
within the high density environment of the forming cluster at that
epoch. 

\begin{figure}
\centering
\epsfxsize=\hsize\epsffile{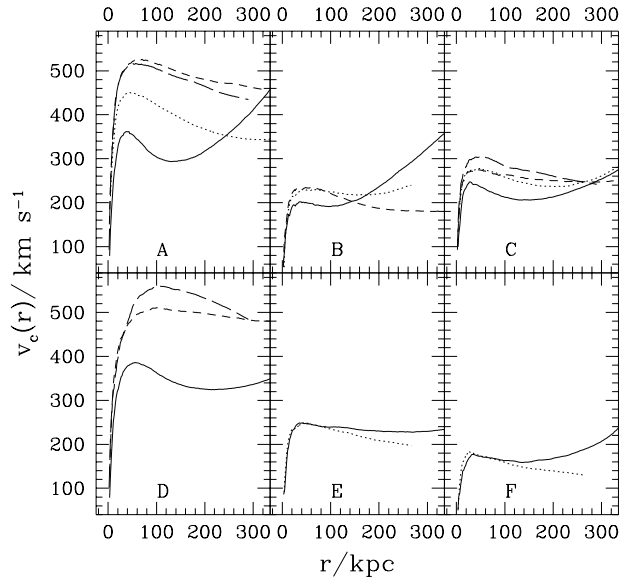}
\caption{
Evolution of the circular velocity profiles for the same
halos as in Figure~\ref{f:rhoevol.1},
plotted using the same line types at the
different redshifts.  The velocities are measured in physical $\kms$.
}
\label{f:vcevol.1}
\end{figure}
The features of $\rho(r)$ are highlighted in the 
corresponding circular velocity profiles $v_c(r)$ shown in 
Figure~\ref{f:vcevol.1}. Note that the location and height of the
peak of $v_c(r)$ have both changed for the {\sl inner} halos 
(upper panels), whilst the profiles
of the {\sl outer} halos E and F, in the lower row, are remarkably stable,
apart from the change in $\rho_{bkg}$.

Halos A and D show the most significant evolution
and 
have quite steep outer slopes. Halo A formed at the intersection of the two
filaments whose collapse originated the cluster and its structure has
been heavily disrupted by the tidal field there, causing it to lose a
huge amount of mass.  Halo D formed in the outer parts of one of the
filaments, and its fate has been similar.
Halos B and C have also evolved, but the changes have not been so
dramatic. The values of $v_{peak}$ have decreased however by $\sim
15$\% between $z=0.5$ and $z=0$.
The remaining halos E and F, that do not show signs of significant evolution,
formed in the periphery,
but at $z=0$ they are moving outwards. 
However the pericenter of F is close to its present
distance ($\sim 1.3\Mpc$). 
That for E is $\sim 600\kpc$, yielding an expected tidal
radius of $\sim 150\kpc$; this agrees very well with the SKID value,
but is 25\% smaller than 
the tidal radius measured from the density profile (directly from
$\rho(r)$, or from $v_c(r)$).
This is due to the anisotropic distribution of mass around E,
that appears stretched along the orbit with a ``trail'' of particles on
both sides.  This tidal debris were contained within its virial radius
at $z=0.5$ and, though stripped during the passage at pericenter and
presently unbound to it, are still moving closely apparently on the
same orbit. The lack of evolution in the density profile of E is then
due to the assumption of spherical symmetry and the contribution from 
unbound particles nearby.

\begin{figure}
\centering
\epsfxsize=\hsize\epsffile{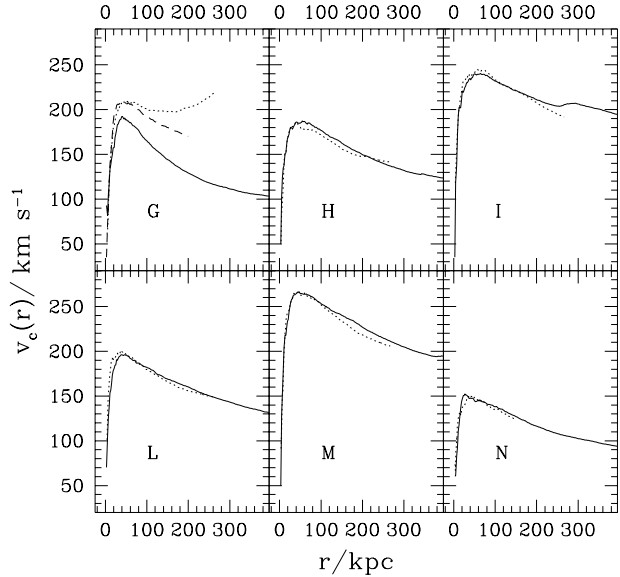}
\caption{
Evolution of the circular velocity profiles of a sample of
peripheral halos after their last major mergers. The dotted and
short--dashed lines correspond to the redshifts $z=0, 0.5, 1$. (At
$z=0$, halo G has a distance from the cluster's center $\sim 4/3
R_{200}$ but was inside the cluster at earlier epochs.}
\label{f:vcevol.2}
\end{figure}
Finally, 
Figure~\ref{f:vcevol.2} shows the corresponding $v_c$ profiles for the 6
peripheral halos considered. Except for case G, 
these halos have never entered the cluster and their profiles 
show no significant differences between $z=0$ and $0.5$. 
Halo G formed in the periphery but has been orbiting through the
cluster since $z\sim 0.5$ and has clearly lost mass.

In summary, of the large cluster halos identified at $z=0$,
only those with $R\mincir R_{200}/2$ have density profiles
significantly different from those they had before the formation of the
cluster. However, one of the large halos
in the outskirts of the cluster has survived a passage (or
more) at pericenter and has evolved considerably too.
In particular, the values of $v_{peak}$ can change: $\sim 10$\% in
three cases, but 20--25\% in two other cases. 

\begin{figure}
\centering
\epsfxsize=\hsize\epsffile{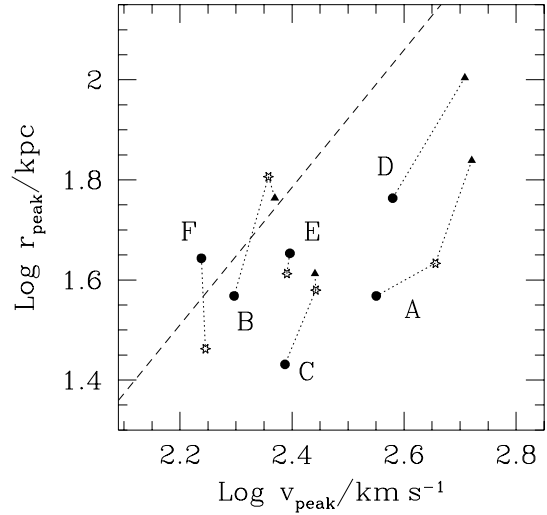}
\caption{
Evolution of the cluster halos of Figure~\ref{f:rhoevol.1} in the plane 
$(v_{peak},r_{peak}$). The positions at $z=0$ are plotted as filled
circles, those at $z=0.5$ as stars (not for D) and at $z=1$ as
triangles (only for A--D). The dashed line is the relation expected for
isolated halos at $z=0$ from Navarro \etal (1996) as in 
Figure~\ref{f:rpk-vpk}.}
\label{f:RPK-VPK}
\end{figure}
\begin{figure}
\centering
\epsfxsize=\hsize\epsffile{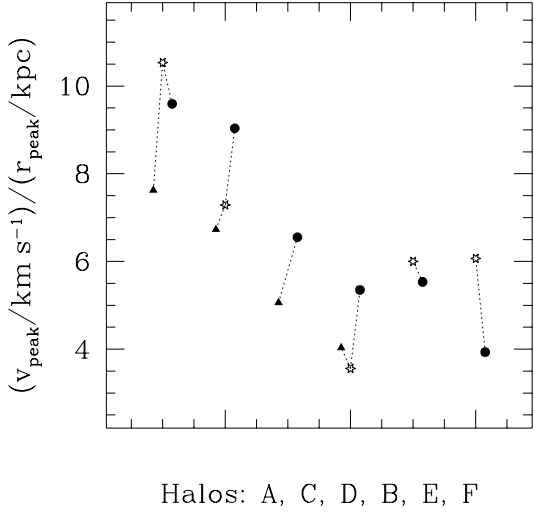}
\caption{
Evolution of ``concentrations'' $v_{peak}/r_{peak}$
for the halos of Figure~\ref{f:RPK-VPK} and the 
points represent the same redshifts as before. The halos
are ordered from left to right according to the overdensity
of the region where they formed: A and B formed at the intersection
of the filaments whose collapse originated the cluster, D along one of 
the filaments and the others in the ``field''.}
\label{f:VPKRPK}
\end{figure}

It is interesting to examine how the halos ``move'' in the plane
($r_{peak}$, $v_{peak}$) of Figure~\ref{f:rpk-vpk}, so as to determine if 
their evolution (under tidal stripping or halo--halo encounters)
is the reason for the systematically lower $r_{pk}$ for the
cluster halos with respect to field halos of same $v_{pk}$.
Figure~\ref{f:RPK-VPK} shows 
the evolution of $r_{pk}$ vs $v_{pk}$ for the 6 cluster halos 
considered previously from $z=1$ (when available) 
through $0.5$ to $z=0$ (from 
triangles to stars to circles in the figure).
\ From these data, it seems that 
the points did not move significantly {\sl away} from the NFW's curve. 
They move {\sl parallel} to the line rather than downwards
and those furthest from the line are also those that formed the
nearest to primordial high--density regions (A and C, that formed
near the cluster's center, but also D, that formed in one of the giant
filaments that merged into it). This could be evidence that halos form
earlier in high--density regions and are thus more concentrated than
those forming in the field.
This impression is strengthened by Figure~\ref{f:VPKRPK}, that shows the
evolution of the ``concentration'' $v_{peak}/r_{peak}$ (the symbols are
as in Figure~\ref{f:RPK-VPK}). The points on the left are for the halos
forming near the center (A and C), those on the right for those
forming in the periphery (B, E, F); the intermediate case D is duely in
the middle.  Although the ratios change, the two groups 
are clearly separated at both epochs with higher values of concentration 
for the former and lower values for the latter group. 
Recently, Lemson \& Kauffmann (1997) have studied correlations
between halo properties and local environment using large $N$--body
simulations. Except for the mass distribution of halos, they do not
find any such correlation. It should be noted however
that their definition of {\sl locality} is based on a scale of 5--$10
h^{-1} \Mpc$, that is still mildly non--linear today. The correlation
we find seems to involve scales of a few Mpc that become non--linear at
early epochs. It will be interesting to examine the significance of our
result with larger samples of cluster halos within simulations of even higher
resolution.

\subsubsection{Analytic fits}

The evolved density profiles of isolated halos in $N$--body
simulations are well described for a large range of masses by the
analytic model of NFW 
(although increasing the numerical resolution causes steeper
inner profiles, Moore \etal (1997); this is not an issue here since 
our sample of halos have similar resolution as those in the NFW simulations):
\be
  {\rho(r)\over \rho_{cr}} = {\delta_c \over (c r/r_{200})(1+c r/r_{200})^2}~,
\ee
\be
  {\mbox{\rm with~~}} \delta_c={200\over 3}{c^3 \over (ln(1+c)-c/(1+c))}~.
\ee
Navarro, Frenk \& White (1997) 
developed an analytic procedure that gives $c$ as a function of
the halo mass $M_{200}$ in any hierarchical cosmological model, based on the
expected redshift of collapse of a density perturbation of mass $M$ in
the Press--Schechter (Press \& Schechter 1974) formalism
Here we address the following two questions: $(i)$ does the NFW
profile and predicted $c$ provide a good description of the profiles
of our {\sl peripheral halos}?  As mentioned previously, this 
question is not trivial since the environment within which these halos evolved is
perturbed by the intense gravitational field of the cluster;
$(ii)$ How much do the tidally stripped {\sl cluster halos} depart from the
NFW predictions ?  For
example, a steeper outer slope may be typical of these halos.
For this reason, we also consider the Hernquist profile (Hernquist
1990; HER in the following):
\be
  {\rho(r)\over \rho_{cr}} = {\delta_c \over (c r/r_{200})(1+c r/r_{200})^3}~,
\ee
\be 
{\mbox{\rm with~~}} \delta_c={400\over 3} c (1+c)^2~. 
\ee
This profile has the same inner form as NFW, but asymptotes to $r^{-4}$ on
large scales instead of $r^{-3}$.

\begin{figure}
\centering
\epsfxsize=\hsize\epsffile{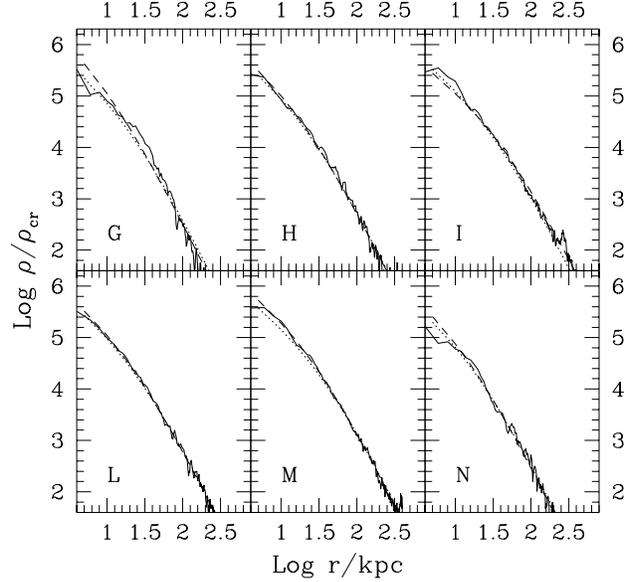}
\caption{
Comparison between the density profiles (at $z=0$) of a sample of 
peripheral halos (solid lines) and the NFW model: the dotted lines are
the expected NFW profiles (from halo virial masses) and the dashed 
lines are fits of the NFW profile to the data.}
\label{f:rhofit3now}
\end{figure}
First we consider the sample of large {\sl peripheral halos} described 
above.
Figure~\ref{f:rhofit3now} shows their density profiles $\rho(r)/\rho_{cr,0}$.
In each panel, the {\sl dotted} and {\sl dashed} lines are NFW profiles 
{\sl predicted} 
using the measured $M_{200}$ according to
the NFW procedure mentioned above, 
and {\sl fit} to the data using a standard $\chi^2$ minimization technique with
$c$ as free parameter (using $r_{200}$ measured from the data);
for the fits, we used the points in the radial range delimited by 
$l_{soft}$ and $r_{200}$.
(Two--parameter fits,
with both $c$ and $r_{200}$ as free parameters, yield very similar
results; we use two--parameter fits for the cluster halos below, since
$r_{200}$ cannot be defined for them.) 
\begin{figure}
\centering
\epsfxsize=\hsize\epsffile{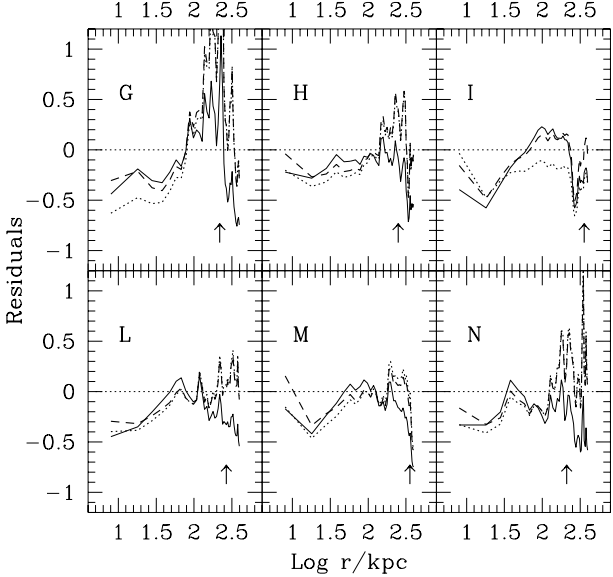}
\caption{
Residuals $(\rho_{NFW}(r)-\rho(r))/\rho(r)$ between the NFW
profiles and density profiles of Figure~\ref{f:rhofit3now} 
(dotted and dashed lines
are again for the expected and fitted NFW profiles).  The residuals of
fits of the Hernquist profile to the data are shown as thin solid
lines. The arrows mark the halo virial radii.(For clarity, 
we show 1 data point every 5).}
\label{f:resid3now}
\end{figure}
We plot 
the residuals between the data and the NFW profiles and Hernquist profiles
in Figure~\ref{f:resid3now}.
Each panel shows the fractional difference
$(\rho_{model}(r)-\rho(r))/\rho(r)$ as a function of $r$, where
$\rho(r)$ are the data and $\rho_{model}(r)$ is the density
corresponding to one of the analytic profiles considered.
With the exception of halo G, the residuals for the NFW profiles are within
$\sim 20$--30\% at distances from the halo centers 
$r\magcir 15\kpc$ 
At smaller distances,
the profiles of our halos are steeper than the NFW curves with 
residuals typically in excess of 30\% for the fits and 
40\% for the predicted curves.
Note that, for the same halo, the different NFW curves have
concentrations that can differ by various amounts, typically from
$c\sim13$--14 (expected) to 16--18 (fits). 
The data do not discriminate significantly 
between NFW and Hernquist's profiles.
Only for halo G, as expected from its steep profile, 
the HER fit fares better, although
it still has positive residuals of $\sim 40$--50\%.

\begin{figure}
\centering
\epsfxsize=\hsize\epsffile{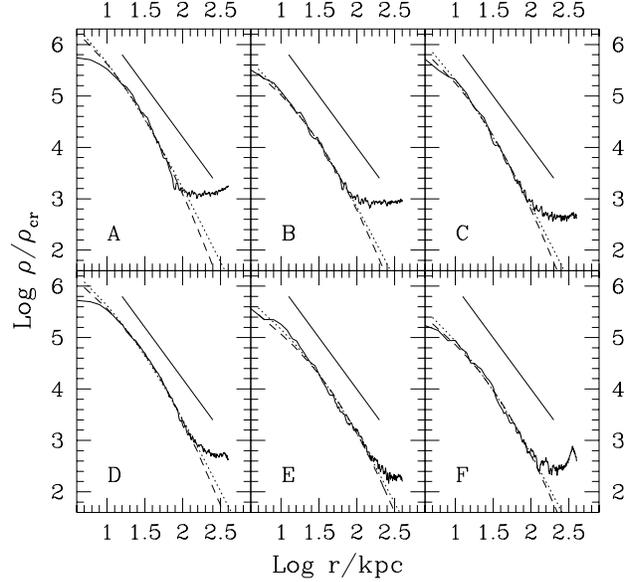}
\caption{
Comparison between the density profiles (at $z=0$) of a sample
of cluster halos (solid lines) and fits of NFW (dotted lines) and Hernquist's
(dashed lines) profiles. An arbitrary isothermal profile is drawn for
comparison (thin solid line).
}
\label{f:rhofit1now}
\end{figure}
\begin{figure}
\centering
\epsfxsize=\hsize\epsffile{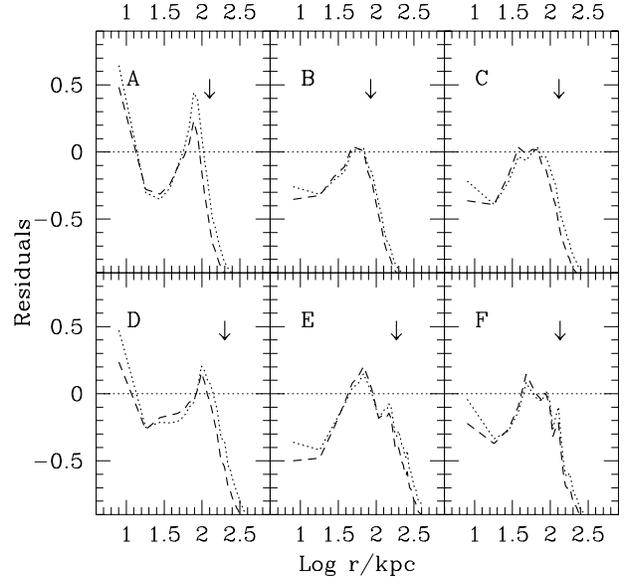}
\caption{
Residuals $(\rho_{FIT}(r)-\rho(r))/\rho(r)$ for the fits
of Figure~\ref{f:rhofit1now}
(NFW, dotted line, and Hernquist, dashed). The arrows mark
the halo tidal radii.
}
\label{f:resid1now}
\end{figure}

Let us now examine the {\sl cluster halos}. Their density profiles (again
solid lines) are shown in Figure~\ref{f:rhofit1now}, together with
NFW (dotted) and HER (dashed) fits (for the latters, we 
use the data in the radial range delimited by $l_{soft}$
and the value where $\rho(r)$ flattens approaching
$\rho_{bkg}$). 
Over the scales of interest, the two fits do
not differ significantly and both underestimate the central
concentrations of the halos.  As before, this can be seen better by
examining the residuals in Figure~\ref{f:resid1now}. 
At intermediate distances from the halo centers ($r\mincir 40\kpc$),
both fits have negative residuals in excess of 
$\sim20$\%, and in excess of $\sim 30$\% on smaller scales. 
At larger distances, up to about 75\% of the halo tidal radii, 
they fare generally
well, with residuals well below 20\%. The exception is halo~A,
that has, as halo G, a particularly steep profile: for this halo, HER is
still an acceptable fit, while NFW has large residuals, $\sim
40$--50\%.  When $r$ approaches $r_{tid}$, the fits depart
rapidly from the flattening $\rho(r)$, with average residuals at
$r_{tid}$ of order of 30\% or more, for NFW, and about 10\% higher for
HER.

As a general remark to conclude this section, if we limit the halo
sample to lie within $r<3/4r_{tid}$, Hernquist's fits are better than NFW
for the profiles of cluster halos (or stripped halos in general),
while the latter model fares well for peripheral halos. Although our
resolution is not good enough to address reliably the issue of halo
concentrations, it is remarkable that our halos
are systematically more concentrated than the NFW model predicts, when from 
the poorer resolution we would expect just the opposite.
In our analysis, the NFW profiles fit to the data have concentrations
about 25\% higher than expected and still, they generally
underestimate the profiles measured at small scales.

\section{Summary and discussion}
\label{s:conc}

We have explored the consequences of increasing the force and mass
resolution within a {\it dark matter only} simulation of a rich galaxy
cluster that forms hierarchically within a cold dark matter
simulation of a closed Universe.  By resimulating regions of interest
using increased resolution, we have attained an unprecedented
view of the internal structure of a massive dark halo. With
approximately 
one million particles within the virial radius and
force softening that is 0.25\% of $R_{200}$, we resolve 150
halos with circular velocity larger than 80 $\kms$ within $R_{200}$
at $z=0$.  Most of these halos have made several orbits within
the cluster and are easily identified as potential minima or density
enhancements above the background.

This work demonstrates that the overmerging of dark matter
substructure within virialised structures can be greatly reduced given
high enough numerical resolution. 
The statement that ``at bottom the problem
[of overmerging] appears to be not numerical but physical: gravitational
 dynamics alone cannot explain the existence of galaxy groups and clusters''
(Weinberg, Katz and Hernquist 1997) is, at the light of these new results 
, completely wrong.

 Overmerging within the cluster
environment is due to the disruption of halos by the global tidal
field and halo--halo encounters 
({\it c.f.} Moore \etal 1996), probably primarily 
within the large dense
halos prior to the formation of the main cluster.  Although our
softening length (5 kpc) 
is a small fraction of the cluster's virial radius,
the rotation curves of the halos peak at radii of $\sim 30$ kpc.
Hence the cores of these halos can be softened to a degree that
effects their evolution. 
This is the primary reason the halos are still being
disrupted; given high enough force and mass resolution, it should be
possible to overcome {\it most} of the overmerging problem within CDM
simulations.

The aim of this paper has been to analyse the properties and dynamics
of the dark matter substructure and we find the following key results:

\begin{itemize}

\item{}  
The orbital distribution of substructure halos is close to that
of an isotropic population of tracers in an isothermal potential;
the median value of apocentric to pericentric distances is
6:1, a value that does not vary with position within the cluster and
is unbiased with respect to the orbits of the smooth particle
background. Circular orbits are rare and 
about 20\% of all our surviving halos
within the cluster will pass within 200 kpc $\equiv 0.1R_{200}$.

\item{} Most dark halos are 
tidally truncated to a value determined by
the (average) density of the cluster at their pericentric positions.  The
approximation of isothermal halo mass distributions orbiting within a
deeper isothermal potential works very well; {\it i.e.}
$r_{tidal}\sim r_{peri}\sigma_{halos}/\sigma_{clus}$. 

\item{} The mass attached to dark matter halos is approximately 13\% of
the entire cluster mass and varies from 0\% within $\sim 200\kpc$ from
the cluster center, to
20\% at its virial radius. This latter value is roughly the expected
value for the mass attached to halos above a circular velocity of $80
\kms$.  Correspondingly, the sizes of halos vary with cluster centric
radius, an effect that may be
observable using gravitational lensing of background galaxies.

\item{} Overmerging within the central regions of dense halos
leads to a final distribution of substructure that is 
antibiased (less centrally concentrated) 
with respect to the global mass distribution. 

\item{} 
The density profiles of a sample of well resolved halos
indicate that those forming in the high--density regions of the collapsing
cluster have higher concentrations than those found in isolated environments.
We show that this is most
probably due to their earlier collapse redshifts rather than the
internal response of the halos to mass loss and heating from tidal
stripping. 
 
\item{} Most of the halos within the cluster and in the cluster
proximity have density profiles that are well fit by NFW profiles
(Navarro, Frenk \& White 1996).
Halos that lose a great deal of mass through tidal stripping have
outer density profiles as steep as $\rho(r) \propto r^{-4}$ (at
$\approx 30\%$ of their virial radius), thus Hernquist profiles
(Hernquist 1990) provide slightly better fits.

\item{} Mergers between halos in the cluster proximity occur with a
frequency of about 5--10\% since z=0.5. In the cluster environment
mergers are rare; not a single merger occurs for halos whose
orbits are contained within $1.6 \Mpc\equiv 80$\%$\,R_{200}$ from 
the cluster center.

\end{itemize}

\section*{Acknowledgments}

We would like to thank R. Carlberg, A. Jenkins, J. Navarro, G. Tormen and 
S. White for interesting discussions. 
SG was supported by the University of Milano. BM is supported by the Royal
Society.  FG acknowledges  a  fellowship from the
 EU network for Galaxy Formation and Evolution.  
We are grateful to Paolo Tozzi and Julio F. Navarro for kindly supplying
fortran routines.
The numerical simulations were carried out at the Denali Arctic Supercomputing
Centre, the IBM SP-2 at the
Cornell Theory Center, and on the Cray T3E at the Pittsburgh
Supercomputing Center with support from an NSF Metacenter grant.

\end{document}